\documentclass[preprint,authoryear,12pt]{elsarticle}

\usepackage{amssymb, epstopdf}
\usepackage{amsmath,color}

\journal{Journal of Statistical Planning and Inference}

\usepackage{subfigure}

\newcommand{\ve}[1]{\mbox{\boldmath ${#1}$}}
\newcommand{\vesub}[2]{\mbox{{\boldmath ${#1}$}$_{#2}$}}

\newcommand{\vess}[3]{\mbox{{\boldmath ${#1}$}$_{#2}^{#3}$}}
\newcommand{\hve}[1]{\hat{\ve{#1}}}

\begin{document}

\def\la{\lambda} \def\f{\noindent} \def\c{\centerline}
\def\ol{\overline} \def\T{{\text{T}}} \def\t{\text}
\def\d{{\text{d}}} \def\om{\omega} \def\Om{\Omega} \def\sub{\subset}
\def\al{\alpha} \def\dt{\delta} \def\ep{\varepsilon} \def\eq{\equiv}
\def\la{\lambda} \def\lg{\langle} \def\rg{\rangle}
\def\e{{{\text{e}}}} \def\R{{\Bbb R}} \def\C{{\Bbb C}}
\def\vp{\varphi} \def\pt{\partial} \def\na{\nabla} \def\Dt{\pi}
\def\pma{\pmatrix} \def\epm{\endpmatrix} \def\Ga{\Gamma}
\def\for{\forall} \def\si{\sigma} \def\wt{\widetilde}
\def\tha{\theta} \def\ga{\gamma} \def\be{\beta}
\def\sm{\sum\limits_{i=1}^n} \def\ti{\tilde}

\begin{frontmatter}


\title{Extended T-process Regression Models}
\author[a]{Zhanfeng Wang}
\address[a]{Department of Statistics and Finance, University of Science and Technology of China, Hefei, China}
\author[b]{Jian Qing Shi\thanks{Correspondence to: Dr J. Q. Shi, School of Mathematics \& Statistics, Newcastle University, UK, j.q.shi@ncl.ac.uk.}}
\address[b]{School of Mathematics and Statistics, Newcastle University, Newcastle, UK}

\author[c]{Youngjo Lee}
\address[c]{Department of Statistics, Seoul National University, Seoul, Korea}

\begin{abstract}

Gaussian process regression (GPR) model has been widely used to fit data when the regression function is unknown  and its nice properties have been well established. In this article, we introduce an extended t-process regression (eTPR) model, a nonlinear model which allows a robust best linear unbiased predictor (BLUP). Owing to its succinct construction, it inherits many attractive properties from the GPR model, such as having closed forms of marginal and predictive distributions to give an explicit form for robust procedures, and easy to cope with large dimensional covariates with an efficient implementation. Properties of the robustness are studied. Simulation studies and real data applications show that the eTPR model gives a robust fit in the presence of outliers in both input and output spaces and has a good performance in prediction, compared with other existed  methods.

\end{abstract}


\end{frontmatter}

\section{Introduction}

Consider a concurrent functional regression model
\begin{equation}
y_{ij}=f_{0i}(\mbox{\boldmath${x}$}_{ij})+\epsilon
_{ij},~~i=1,...,m,j=1,...,n_i,  \label{true}
\end{equation}
where {  $m$ is number of functional curves, $n_i$ is number of observed data points for the $i$-th curve,} $f_{0i}(\mbox{\boldmath${x}$}_{ij})$ is the value of unknown function $%
f_{0i}(\cdot )$ at the $p\times 1$ observed covariate $%
\mbox{{\boldmath
${x}$}$_{ij}$}\in \mathcal{X}=R^{p}$ and $\epsilon _{ij}$ is an error term.
To fit unknown functions $f_{0i},$ we may consider a process regression
model
\begin{equation}
y_i(\mbox{\boldmath ${x}$})=f_i(\mbox{\boldmath${x}$})+\epsilon_i(%
\mbox{\boldmath
${x}$}),~~i=1,...,m,  \label{assumed}
\end{equation}
where $f_i(\mbox{\boldmath${x}$})$ is a random function and $\epsilon_i(%
\mbox{\boldmath
${x}$})$ is an error process for $\mbox{\boldmath ${x}$}\in \mathcal{X}$. 
A
GPR model assumes a Gaussian process (GP) for the random function $f_i(\cdot
)$. It has been widely used to fit data when the regression function is
unknown: for detailed descriptions see \cite{r12}
, and \cite{r16} 
and references therein.
GPR has many good features, for example, it can model nonlinear relationship
nonparametrically between a response and a set of large dimensional
covariates with efficient implementation procedure. In this paper we
introduce an eTPR model and investigate advantages in using an extended
t-process (ETP).

BLUP procedures in linear mixed model are widely used \citep{r13} and 
extended to Poisson-gamma models \citep{r8} and Tweedie models
\citep{r11}. Efficient BLUP algorithms have been developed for
genetics data \citep{r24} and spatial data \citep{r4}. In this paper, we show that BLUP procedures can be extended to GPR
models. However, GPR does not give a robust inference against outliers in
output space ($y_{ij} $). Locally weighted scatterplot smoothing (LOESS) method \citep{r3} has been
developed for a robust inference against such outliers. However, it requires
fairly large densely sampled data set to produce good models and does not
produce a regression function that is easily represented by a mathematical
formula. For models with many covariates, it is inevitable to have sparsely
sampled regions. \cite{r18} showed that the GPR model tends
to give an overfit for data points in the sparsely sampled regions (outliers
in the input space, $\mbox{\boldmath${x}$}_{ij}$). Thus, it is important to
develop a method which produces robust fits for sparsely sampled regions as
well as densely sampled regions.  \cite{r18} proposed to use
a heavy-tailed process. However, their copula method does not lead to a
close form for prediction of $f_i(\mbox{\boldmath${x}$}).$ As an alternative
to generate a heavy-tailed process, various forms of student $t$-process
have been developed: see for example \cite{r21}, \cite{r23}, \cite{r1} and \cite{r20}.
However, \cite{r15} noted that the $t$-distribution is not
closed under addition to maintain nice properties in Gaussian models.

In this paper, we use the idea in \cite{r9} for double hierarchical generalized linear models  to extend Gaussian process regression to a t-process regression model. This leads to a specific
eTPR model which retains almost all favorable
properties of GPR models, for example marginal and predictive distributions
are in closed forms. The proposed eTPR model includes  \cite{r15}'s t-process model as a special case. We study the extended t-process and its robust
BLUP procedure against outliers in both input and output spaces.
 Properties of the robustness and consistency  are also investigated. In
addition, we want to emphasize  the following two points: (1) the proposed eTPR
model provides a very general model.
Particularly, when $m>1$, the parameter associated with the degrees of freedom in $t$ process can be estimated.
(2) We correct the statement made in
\cite{r15} and clarify which covariance functions can result in
different predictions in GPR and eTPR (see the detailed discussion in
Section 3.2).

The remainder of the paper is organized as follows. Section 2 presents an ETP and its
properties. Section 3 proposes eTPR models and discuss as the inference and
implementation procedures. Robustness properties and information consistency
are shown in Section 4. Numerical studies and
real examples are presented in Section 5, followed by concluding remarks in Section 6.
All the proofs are in Appendix.

\section{Extended $t$-process}

\begin{figure}[h]
\begin{center}
\includegraphics[height = 0.7\textwidth,width=0.98\textwidth]{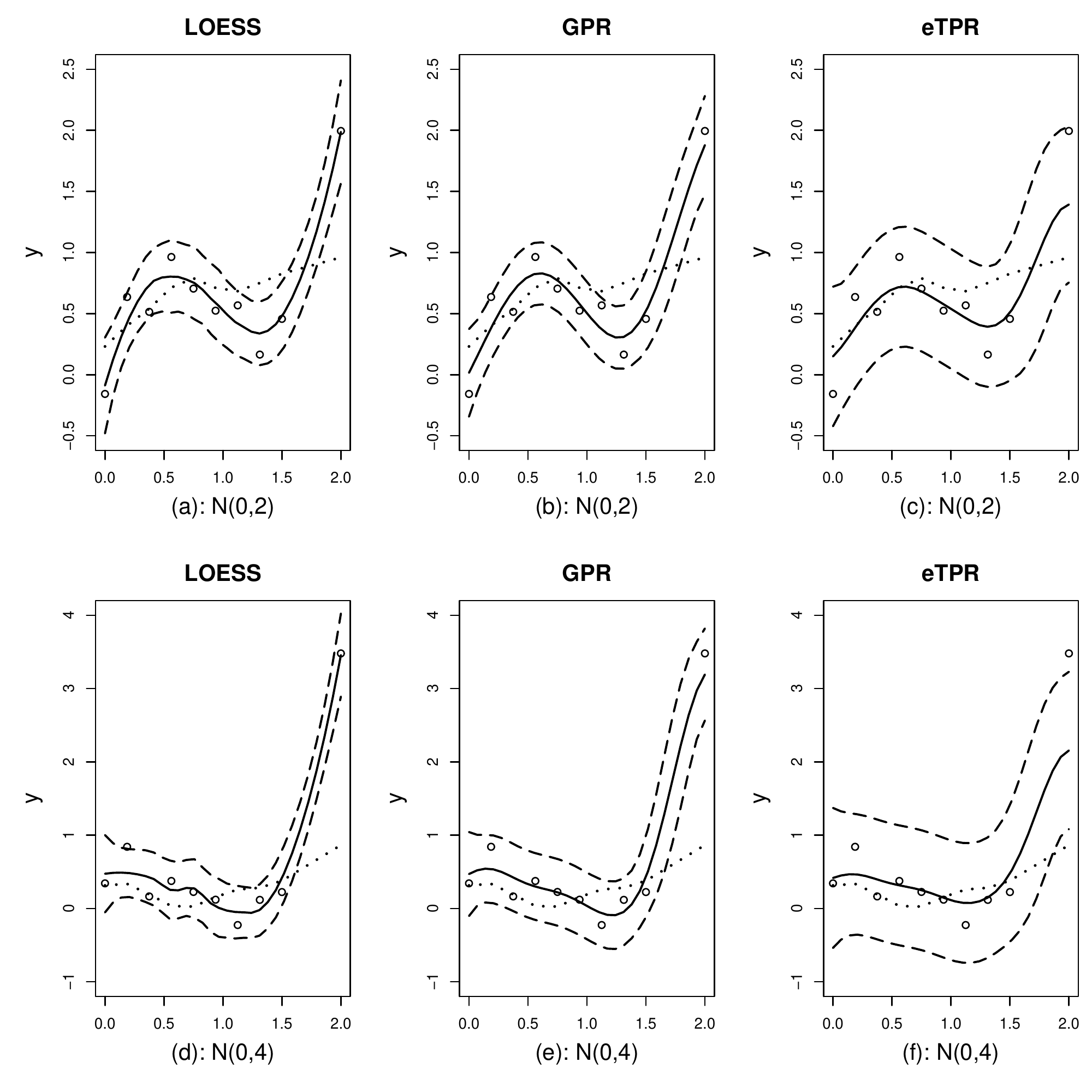}
\end{center}
\caption{Predictions in the presence of outlier at data point 2.0 which is disturbed
by additional error generated from $N(0,\protect\sigma^2)$, where circles represent the observed
data, dotted line is the true function, and solid and dashed lines
stand for predicted curves and their 95\% point-wise confidence intervals from the LOESS,
GPR and eTPR methods respectively. }
\label{fig1}
\end{figure}

As a motivating example, we generated two data sets with $m=1$ and sample
size of $n_1=10$ where $x_{1j}$'s are evenly spaced in [0,~1.5] for the
first $n_1-1$ data points and the remaining point is at 2.0. Thus, this point is a sparse one, meaning it is far away from the other data
points in the input space. In addition, we make the data point 2.0 to be an
outlier in output space by adding an extra error from either $N(0,2)$ or $%
N(0,4)$. To fit the simulated data, covariance kernel takes a combination of
squared exponential kernel and Mat$\acute{e}$rn kernel (see the detailed discussion in
Section 3.2). Prediction curves,  the observed data, the true
function and their 95\%
point-wise confidence intervals, are plotted in Figure \ref
{fig1}. It shows that the predictions obtained from LOESS and GPR are very similar
but the  predictions from eTPR shrinks heavily in the area near the data point 2.0, i.e.
selective shrinkage occurs. {  Another example with an outlier in the middle of the interval is presented in Figure D.6 and shows a similar phenomenon.   } This shows the robustness of eTPR. However, in some other numerical studies, we found that the difference between GPR and eTPR is ignorable. This motivates us to study the eTPR models carefully and comprehensively in both theory and implementation.

Denote the observed data set by $\mbox{{\boldmath
${{\mathcal{D}}}$}$_{n}$}=\{\mbox{{\boldmath ${X}$}$_{i}$},%
\mbox{{\boldmath
${y}$}$_{i}$},i=1,...,m\}$ where $\mbox{{\boldmath${y}$}$_{i}$}%
=(y_{i1},...,y_{in_{i}})^{T}$ and $\mbox{{\boldmath${X}$}$_{i}$}=(%
\mbox
{{\boldmath${x}$}$_{i1}$},...,\mbox{{\boldmath${x}$}$_{in_i}$})^{T}$. For a
random component $f_{i}(\mbox{\boldmath ${u}$})$ at a new point $%
\mbox{\boldmath ${u}$}\in \mathcal{X}$, the best unbiased predictor is $%
E(f_{i}(\mbox{\boldmath
${u}$})|\mbox{{\boldmath
${{\mathcal{D}}}$}$_{n}$}).$ It is called a BLUP if it is linear in \{$%
\mbox{{\boldmath
${y}$}$_{i}$},i=1,...,m$\}. Its standard error can be estimated with $%
Var(f_{i}(\mbox{\boldmath ${u}$})|\mbox{{\boldmath
${{\mathcal{D}}}$}$_{n}$}).$ To have an efficient implementation procedure,
it is useful to have explicit forms for the predictive distribution $p(f_{i}(%
\mbox{\boldmath ${u}$})|\mbox{{\boldmath
${{\mathcal{D}}}$}$_{n}$}),$ and the first two moments $E(f_{i}(%
\mbox{\boldmath
${u}$})|\mbox{{\boldmath
${{\mathcal{D}}}$}$_{n}$})$ and $Var(f_{i}(\mbox{\boldmath ${u}$})|%
\mbox{{\boldmath
${{\mathcal{D}}}$}$_{n}$})$.

Let $f$ be a real-valued random function such that $f: {\mathcal{X}}%
\rightarrow R$. In this paper, we extend a Gaussian process to a t-process using the idea in   Lee and Nelder (2006):
\begin{equation*}
f|r\sim GP(h,rk),~~~~r\sim \mathrm{IG}(\nu ,\omega ),
\end{equation*}
where $GP(h,rk)$ stands for a GP with mean function $h$ and covariance
function $rk$, and $\mathrm{IG}(\nu ,\omega )$ stands for an inverse gamma
distribution.
Then, $f$ follows an ETP $f\sim
ETP(\nu ,\omega ,h,k),$ implying that for any collection of points $%
\mbox{\boldmath
${X}$}=(\mbox{{\boldmath${x}$}$_{1}$},...,\mbox{{\boldmath${x}$}$_{n}$})^{T},%
\mbox{{\boldmath${x}$}$_{i}$}\in {\mathcal{X}}$, we have
\begin{equation*}
\mbox{{\boldmath${f}$}$_{n}$}=f(\mbox{{\boldmath ${X}$}$$})=(f(%
\mbox{{\boldmath${x}$}$_{1}$}),...,f(\mbox {{\boldmath${x}$}$_{n}$}%
))^{T}\sim EMTD(\nu ,\omega ,\mbox{{\boldmath
${h}$}$_{n}$},\mbox{{\boldmath${K}$}$_{n}$}),
\end{equation*}
meaning that $%
\mbox{{\boldmath
${f}$}$_{n}$}$ has an extended multivariate $t$-distribution (EMTD) with the
density function,
\begin{equation*}
p(z)=|2\pi \omega \mbox{{\boldmath ${K}$}$_{n}$}|^{-1/2}\frac{\Gamma
(n/2+\nu )}{\Gamma (\nu )}\left( 1+\frac{(z-\mbox{{\boldmath ${h}$}$_{n}$}%
)^{T}\mbox{{\boldmath ${K}$}$_{n}^{-1}$}(z-\mbox{{\boldmath ${h}$}$_{n}$})}{%
2\omega }\right) ^{-(n/2+\nu )},
\end{equation*}
$\mbox{{\boldmath${h}$}$_{n}$}=(h(\mbox{{\boldmath${x}$}$_{1}$}),...,h(%
\mbox{{\boldmath${x}$}$_{n}$}))^{T}$, $\mbox{{\boldmath
${K}$}$_{n}$}=(k_{ij})_{n\times n}$ and $k_{ij}=k(%
\mbox{{\boldmath
${x}$}$_{i}$},\mbox{{\boldmath${x}$}$_{j}$})$ for some mean function $%
h(\cdot ): {\mathcal{X}}\rightarrow R$ and covariance kernel $k(\cdot ,\cdot
): {\mathcal{X}}\times {\mathcal{X}}\rightarrow R.$

It follows that at any collection of finite points ETP has an analytically
representable EMTD density being similar to GP having multivariate normal
density. Note that $E(\mbox{{\boldmath${f}$}$_{n}$})=%
\mbox
{{\boldmath${h}$}$_{n}$}$ is defined when $\nu >1/2$ and $Cov(%
\mbox
{{\boldmath${f}$}$_{n}$})=\omega \mbox{{\boldmath${K}$}$_{n}$}/(\nu -1)$ is
defined when $\nu >1$. When $\nu =\omega =\alpha /2$, $%
\mbox{{\boldmath
${f}$}$_{n}$}$ becomes the multivariate $t$-distribution of \cite{r6}. When $\nu =\alpha /2$ and $\omega =\beta /2$, $%
\mbox{{\boldmath
${f}$}$_{n}$}$ becomes the generalized multivariate $t$-distribution of
\cite{r2}. For $f\sim ETP(\nu ,\omega ,0,k)$ it
easily obtains that $E(f(\mbox{\boldmath${x}$}))=0,~~Var(f(%
\mbox{\boldmath${x}$}))={\omega }k(\mbox{\boldmath${x}$},\mbox{%
\boldmath${x}$})/{(\nu -1)}$, and
\begin{align}
& Skewness(f(\mbox{\boldmath${x}$}))=\frac{E(f^{3}(\mbox{\boldmath${x}$}))}{%
(E(f^{2}(\mbox{\boldmath${x}$})))^{3/2}}=0,  \notag \\
& Kurtosis(f(\mbox{\boldmath${x}$}))=\frac{E(f^{4}(\mbox{\boldmath${x}$}))}{%
(E(f^{2}(\mbox{\boldmath${x}$})))^{2}}=\frac{3}{\nu -2}+3\geq 3\text{ when }%
\nu >2.  \notag
\end{align}
Thus, we may say that the $ETP(\nu ,\omega ,0,k)$ has a heavier tail than
the $GP(0,k)$.

\vskip10pt \noindent \textbf{Proposition 1} \textit{Let $f\sim ETP(\nu
,\omega ,h,k).$ }

\begin{itemize}
\item[(i)]  \textit{When $\omega /\nu \rightarrow \lambda $ as $\nu
\rightarrow \infty $, we have $\lim_{\nu \rightarrow \infty }ETP(\nu ,\omega
,h,k)=GP(h,\lambda k).$}

\item[(ii)]  \textit{Let $\mbox{\boldmath${Z}$}\in \mathcal{X}$ be a p$%
\times 1$ random vector such that $\mbox{\boldmath${Z}$}\sim EMTD(\nu
,\omega ,\mbox{{\boldmath${\mu}$}$_{z}$},\mbox{{\boldmath${\Sigma}$}$_{z}$})$%
. For a linear system $f(\mbox{\boldmath ${x}$})=%
\mbox{{\boldmath
${x}$}$^{T}$}\mbox{\boldmath ${Z}$}$ with $\mbox{\boldmath ${x }$}\in
\mathcal{X}$, we have $f\sim ETP(\nu ,\omega ,h,k)$ with $h(%
\mbox{\boldmath
${x}$})=\mbox{{\boldmath ${x}$}$^{T}$}\mbox{{\boldmath
${\mu}$}$_{z}$}$ and $k(\mbox{{\boldmath ${x}$}$_{i}$},%
\mbox{{\boldmath
${x}$}$_{j}$})=\mbox{{\boldmath ${x}$}$_{i}^{T}$}%
\mbox{{\boldmath
${\Sigma}$}$_{z}$}\mbox{{\boldmath ${x}$}$_{j}$}$ }.

\item[(iii)]  \textit{\ Let $\mbox{\boldmath ${u}$}\in \mathcal{X}$ be a new
data point and $\mbox{{\boldmath ${k}$}$_{ u}$}=(k(\mbox{\boldmath${u}$},%
\mbox{{\boldmath${x}$}$_{1}$}),...,k(\mbox{\boldmath${u}$},%
\mbox{{\boldmath${x}$}$_{n}$}))^{T}.$ Then, $f|%
\mbox{{\boldmath
${f}$}$_{n}$}\sim ETP(\nu ^{\ast },\omega ^{\ast },h^{\ast },k^{\ast })$
with $\nu ^{\ast }=\nu +n/2,$ $\omega ^{\ast }=\omega +n/2$,
\begin{align}
& h^{\ast }(\mbox{\boldmath${u}$})=\mbox {{\boldmath${k}$}$_{ u}^{T}$}%
\mbox{{\boldmath${K}$}$_{n}^{{-1}}$}(\mbox{{\boldmath${f}$}$_{n}$}-%
\mbox{{\boldmath${h}$}$_{n}$})+h(\mbox {\boldmath${u}$}),  \notag \\
& \mathit{k^{\ast }(\mbox{\boldmath ${u}$},\mbox{\boldmath ${v}$})}=\frac{{%
2\omega +(\mbox {{\boldmath${f}$}$_{n}$}-\mbox{{\boldmath${h}$}$_{n}$})^{T}%
\mbox {{\boldmath${K}$}$_{n}^{{-1}}$}(\mbox{{\boldmath${f}$}$_{n}$}-\mbox
{{\boldmath${h}$}$_{n}$})}}{{2\omega +n}}\ \left( k(\mbox{\boldmath
${u}$},\mbox{\boldmath ${v}$})-\mbox{{\boldmath
${k}$}$_{ u}^{T}$}\mbox{{\boldmath${K}$}$_{n}^{{-1}}$}\mbox{{%
\boldmath${k}$}$_{ v}$}\right) ,  \notag
\end{align}
for $\mbox{\boldmath ${v }$}\in \mathcal{X}$.}
\end{itemize}

Even if the mean and covariance functions of $f\sim ETP(\nu ,\omega ,h,k)$
can not be defined when $\nu <0.5$, from Proposition 1(iii), the mean and
covariance functions of the conditional process $f|%
\mbox{{\boldmath
${f}$}$_{n}$}$ do always exist if $n\geq 2$. Also from Proposition 1(iii),
the conditional process $ETP(\nu ^{\ast },\omega ^{\ast },h^{\ast },k^{\ast
})$ converges to a GP, as either $\nu $ or $n$ tends to $\infty $. Thus, if
the sample size $n$ is large enough, the ETP behaves like a GP.

For a new point $\mbox{\boldmath ${u}$}$, we have $f(\mbox{\boldmath ${u}$})|%
\mbox{{\boldmath ${f}$}$_{n}$}\sim EMTD(\nu ^{\ast },\omega ^{\ast },h^{\ast
}(\mbox{\boldmath ${u}$}),k^{\ast }(\mbox{\boldmath ${u}$},%
\mbox{\boldmath
${u}$}))$, where
\begin{align}
& h^{\ast }(\mbox{\boldmath ${u}$})=E(f(\mbox{\boldmath ${u}$})|%
\mbox{{\boldmath ${f}$}$_{n}$})=\mbox {{\boldmath${k}$}$_{ u}^{T}$}%
\mbox{{\boldmath
${K}$}$_{n}^{{-1}}$}(\mbox{{\boldmath${f}$}$_{n}$}-%
\mbox{{\boldmath
${h}$}$_{n}$})+h(\mbox{\boldmath ${u}$}),  \notag \\
& Var(f(\mbox{\boldmath ${u}$})|\mbox{{\boldmath ${f}$}$_{n}$})=\frac{\omega
^{\ast }}{\nu ^{\ast }-1}k^{\ast }(\mbox{\boldmath ${u}$},%
\mbox{\boldmath
${u}$})=s~\{k(\mbox{\boldmath ${u}$},\mbox{\boldmath ${u}$})-%
\mbox{{\boldmath
${k}$}$_{ u}^{T}$}\mbox{{\boldmath${K}$}$_{n}^{{-1}}$}\mbox{{%
\boldmath${k}$}$_{ u}$}\},  \notag
\end{align}
and $s=({2\omega +(\mbox{{\boldmath${f}$}$_{n}$}-\mbox{{%
\boldmath${h}$}$_{n}$})^{T}\mbox{{\boldmath${K}$}$_{n}^{{-1}}$}(%
\mbox{{\boldmath${f}$}$_{n}$}-\mbox{{\boldmath${h}$}$_{n}$})})/({2\nu +n-2})$%
. Note that from Lemma 2(iv) in Appendix A, $s=E(r|\mbox{{\boldmath${f}$}$_{n}$})$.

Under various combinations of $\nu $ and $\omega $, the ETP generates
various $t$-processes proposed in the literature. For example, $ETP(\alpha
/2,\alpha /2-1,h,k)$ is the $t$-process of \cite{r15}. They
showed that if covariance function $\mbox{\boldmath ${\Sigma }$}$ follows an
inverse Wishart process with parameter $\alpha =2\nu $ and kernel function $%
k $, and $f|\mbox{\boldmath ${\Sigma }$}\sim GP(h,(\alpha -2)%
\mbox{\boldmath
${\Sigma }$})$, then $f$ has an extended $t$-process $ETP(\alpha /2,\alpha
/2-1,h,k)$. $ETP(\alpha /2,\alpha /2,h,k)$ is the Student's $t$-process of
 \cite{r12} and $ETP(\nu ,1/2,h,k)$ is the model discussed in  \cite{r23}.

\section{eTPR models}

For the process regression model (\ref{assumed}), this paper assumes that
$(f_i,\epsilon_i),i=1,...,m$, are independent, and $f_i$ and $\epsilon_i$
have a joint ETP process,
\begin{equation}
\left(
\begin{array}{c}
f_i \\
\epsilon_i
\end{array}
\right) \sim ETP\left( \nu ,\omega ,\left(
\begin{array}{c}
h_i \\
0
\end{array}
\right) ,\left(
\begin{array}{cc}
k_i & 0 \\
0 & k_{\epsilon}
\end{array}
\right) \right) ,  \label{assum1}
\end{equation}
where $h_i$ and $k_i$ are mean and kernel functions, $k_{\epsilon}(%
\mbox{\boldmath ${u}$},\mbox{\boldmath ${v}$})=\phi I(\mbox{\boldmath ${u}$}=%
\mbox{\boldmath ${v}$})$ and $I(\cdot )$ is an indicator function. We can
construct the ETP (\ref{assum1}) hierarchically as
\begin{equation*}
\left(
\begin{array}{c}
f_i \\
\epsilon_i
\end{array}
\right) \Big|r_i\sim GP\left( \left(
\begin{array}{c}
h_i \\
0
\end{array}
\right) ,r_i\left(
\begin{array}{cc}
k_i & 0 \\
0 & k_{\epsilon}
\end{array}
\right) \right) {\mbox{~ and~ }}r_i\sim \mathrm{{IG}(\nu ,\omega )},
\end{equation*}
which implies that $f_i+\epsilon_i |r_i\sim GP(h_i,r_i(k_i+k_{\epsilon}))$ and $%
r_i\sim \mathrm{{IG}(\nu ,\omega )}$
results in $y_i\sim ETP(\nu ,\omega ,h_i,k_i+k_{\epsilon}).$ We call the above model as an extended t-process regression model (eTPR). Hence, additivity
property of the GPR and many other properties hold conditionally and
marginally for the eTPR. When $r_i=1$, the eTPR model becomes a GPR model.
Without loss of generality, let $n_1=\cdots=n_m=n$. For observed data $%
\mbox{{\boldmath
${{\mathcal{D}}}$}$_{n}$}=\{\mbox{{\boldmath ${X}$}$_{i}$},%
\mbox{{\boldmath
${y}$}$_{i}$},i=1,...,m\}$ with $\mbox{{\boldmath${y}$}$_{i}$}%
=(y_{i1},...,y_{in})^{T}$ and $\mbox{{\boldmath${X}$}$_{i}$}=(%
\mbox
{{\boldmath${x}$}$_{i1}$},...,\mbox{{\boldmath${x}$}$_{in}$})^{T}$, the model can be expressed as
\begin{align}
& f_i(\mbox{{\boldmath ${X}$}$_{i}$})|\mbox{{\boldmath ${X}$}$_{i}$}\sim
EMTD(\nu ,\omega ,\mbox{{\boldmath
${h}$}$_{in}$},\mbox{{\boldmath ${K}$}$_{in}$}),  \notag \\
& \mbox{{\boldmath
${y}$}$_{i}$}|f_i,\mbox{{\boldmath ${X}$}$_{i}$}\sim EMTD(\nu ,\omega ,f_i(%
\mbox{{\boldmath ${X}$}$_{i}$}),\phi \mbox{{\boldmath ${I}$}$_{n}$}),  \notag
\\
& \mbox{{\boldmath ${y}$}$_{i}$}|\mbox{{\boldmath ${X}$}$_{i}$}\sim EMTD(\nu
,\omega ,\mbox{{\boldmath ${h}$}$_{in}$},%
\mbox{{\boldmath
${\Sigma}$}$_{in}$}),  \notag
\end{align}
where $\mbox{{\boldmath${h}$}$_{in}$}=(h_i(\mbox{{\boldmath${x}$}$_{i1}$}%
),...,h_i(\mbox{{\boldmath${x}$}$_{in}$}))^{T}$, $%
\mbox{{\boldmath
${K}$}$_{in}$}=(k_{ijl})_{n\times n}$, $k_{ijl}=k_i(%
\mbox{{\boldmath
${x}$}$_{ij}$},\mbox{{\boldmath${x}$}$_{il}$})$, and $\mbox{{\boldmath
${\Sigma}$}$_{in}$}=\mbox{{\boldmath
${K}$}$_{in}$}+\phi \mbox{{\boldmath ${I}$}$_{n}$}$.

Consider a linear mixed model
\begin{equation*}
y_{1j}=\mbox{{\boldmath ${w}$}$_{j}^{T}$}\mbox{\boldmath ${\delta }$}+%
\mbox{{\boldmath ${v}$}$_{j}^{T}$}\mbox{\boldmath ${b}$}+\epsilon
_{1j},~~j=1,...,n,
\end{equation*}
where \mbox{{\boldmath ${w}$}$_{j}$} is the design matrix for fixed effects $%
\mbox{\boldmath ${\delta}$}$, $\mbox{{\boldmath ${v}$}$_{j}$}$ is the design
matrix for random effect $\mbox{\boldmath ${b}$}\thicksim N(0,\theta %
\mbox{{\boldmath ${I}$}$_{p}$})$ and $\epsilon _{1j}\thicksim N(0,\phi )$ is
a white noise. Suppose that $\mbox{{\boldmath ${X}$}$_{1}$}=(%
\mbox{{\boldmath
${W}$}$_{n}$},\mbox{{\boldmath ${V}$}$_{n}$})$, $\ f_1(%
\mbox{{\boldmath
${X}$}$_{1}$})=\mbox{{\boldmath ${W}$}$_{n}^{T}$}\mbox{\boldmath ${\delta}$}+%
\mbox{{\boldmath ${V}$}$_{n}^{T}$}\mbox{\boldmath ${b
}$}$, $\mbox{{\boldmath ${h}$}$_{1n}$}=\mbox{{\boldmath ${W}$}$_{n}^{T}$}%
\mbox{\boldmath ${\delta}$}$ and $\mbox{{\boldmath ${K}$}$_{1n}$}=\theta %
\mbox{{\boldmath ${V}$}$_{n}$}\mbox{{\boldmath ${V}$}$_{n}^{T}$}$ with $%
\mbox{{\boldmath ${W}$}$_{n}$}=(\mbox{{\boldmath ${w}$}$_{1}$},...,%
\mbox{{\boldmath ${w}$}$_{n}$})^{T}$ and $\mbox{{\boldmath ${V}$}$_{n}$}=(%
\mbox{{\boldmath ${v}$}$_{1}$},...,\mbox{{\boldmath ${v}$}$_{n}$})^{T}$.
Then, the linear mixed model becomes the functional regression model with
\begin{align}
& f_1(\mbox{{\boldmath ${X}$}$_{1}$})|\mbox{{\boldmath ${X}$}$_{1}$}=%
\mbox{{\boldmath ${W}$}$_{n}^{T}$}\mbox{\boldmath ${\delta}$}+%
\mbox{{\boldmath ${V}$}$_{n}^{T}$}\mbox{\boldmath ${b}$}|%
\mbox{{\boldmath
${X}$}$_{1}$}\sim N(\mbox{{\boldmath ${W}$}$_{n}^{T}$}%
\mbox{\boldmath ${\delta
}$},\mbox{{\boldmath ${K}$}$_{1n}$}),  \notag \\
& \mbox{{\boldmath
${y}$}$_{1}$}|f_1,\mbox{{\boldmath ${X}$}$_{1}$}=%
\mbox{{\boldmath
${y}$}$_{1}$}|\mbox{\boldmath ${b}$},\mbox{{\boldmath ${X}$}$_{1}$}\sim N(%
\mbox{{\boldmath ${W}$}$_{n}^{T}$}\mbox{\boldmath ${\delta}$}+%
\mbox{{\boldmath ${V}$}$_{n}^{T}$}\mbox{\boldmath ${b
}$},\phi \mbox{{\boldmath ${I}$}$_{n}$}),  \notag \\
& \mbox{{\boldmath ${y}$}$_{1}$}|\mbox{{\boldmath ${X}$}$_{1}$}\sim N(%
\mbox{{\boldmath ${W}$}$_{n}^{T}$}\mbox{\boldmath ${\delta }$},%
\mbox{{\boldmath
${\Sigma}$}$_{1n}$}).  \notag
\end{align}
This shows that the eTPR model extends the conventional normal linear mixed
models to a nonlinear concurrent functional regression. Contrary to LOESS, this also
shows that the eTPR method can produce a regression function, easily
represented by a mathematical formula.

\subsection{Parameter estimation }

{So far we have assumed that the covariance kernel $k_i(\cdot ,\cdot )$ is
given. To fit the eTPR model, we need to choose $k_i(\cdot ,\cdot )$. A way
is to estimate the covariance kernel nonparametrically; see e.g.  \cite{r5}. However, this method is very difficult to be applied
to problems with multivariate covariates. Thus, we choose a covariance
kernel from a covariance function family such as a squared exponential kernel and
Mat\'{e}rn class kernel.}

Let $\mbox{\boldmath ${\beta}$}=(\phi ,\mbox{\boldmath ${\theta}$}_1,...,%
\mbox{\boldmath ${\theta}$}_m)$, where $\phi $ is a parameter for $\epsilon (%
\mbox{\boldmath ${x}$})$ and $\mbox{\boldmath ${\theta}$}_i$ are those for $%
f_i(\mbox{{\boldmath ${x}$}})$ (parameters involved in the kernel $k_i$), $i=1,...,m$.
Because $\mbox{{\boldmath
${y}$}$_{i}$}|\mbox{{\boldmath ${X}$}$_{i}$}\sim EMTD(\nu ,\omega,0,%
\mbox{{\boldmath
${\Sigma}$}$_{in}$})$, the maximum likelihood (ML) estimator  for $\mbox{\boldmath ${\beta}$}$ can be
obtained by solving
\begin{equation}
\sum_{i=1}^m\frac{\partial \log p{_{\beta }(\mbox{{\boldmath
${y}$}$_{i}$}|\mbox{{\boldmath ${X}$}$_{i}$})}}{\partial\beta_k}=\frac{1}{2}%
\sum_{i=1}^m Tr\left( \Big(s_{1i}\mbox{\boldmath
${\alpha}_i$}\mbox{{\boldmath
${\alpha}_i$}$^{T}$}-\mbox{{\boldmath ${{\Sigma}}$}$_{in}^{-1}$}\Big)%
\frac{\partial \mbox{{\boldmath ${{\Sigma}}$}$_{in}$}}{\partial \beta_k%
}\right) =0,  \label{score-beta-1}
\end{equation}
where $\beta_k$ is the $k$th element of $\mbox{\boldmath ${\beta}$}$, $%
\mbox{\boldmath ${\alpha}_i$}=%
\mbox{{\boldmath
${{\Sigma}}$}$_{in}^{-1}$}\mbox{{\boldmath ${y}$}$_{i}$}$, $s_{1i}=({%
n+2\nu })/({2\omega+\mbox{{\boldmath ${y}$}$_{i}^{T}$}%
\mbox{{\boldmath
${{\Sigma}}$}$_{in}^{-1}$}\mbox{{\boldmath ${y}$}$_{i}$}})$, and
\begin{equation}
p_{\beta }(\mbox{{\boldmath ${y}$}$_{i}$}|\mbox{{\boldmath ${X}$}$_{i}$}%
)=|2\pi \omega\mbox{{\boldmath
${{\Sigma}}$}$_{in}$}|^{-1/2}\frac{\Gamma (n/2+\nu )}{\Gamma (\nu )}%
\left( 1+\frac{\mbox{{\boldmath ${y}$}$_{i}^{T}$}%
\mbox{{\boldmath
${\Sigma}$}$_{in}^{-1}$}\mbox{{\boldmath
${y}$}$_{i}$}}{2\omega}\right) ^{-(n/2+\nu )}.  \label{margin}
\end{equation}
Score equations for GPR models are the ML estimating equations above with $%
\nu =\infty $ and $s_{1i}=1$. Thus, a parameter estimation for the eTPR models can be obtained by a modification of the existing
procedures for the GPR models.

From the hierarchical construction of ETP with $m=1$, there is only one
single random effect $r_{1}$, so that $r_{1}$ is not estimable, confounded
with parameters in covariance matrix. This means that $\nu $ and $\omega $
are not estimable. Following  \cite{r9}, we set $\omega =\nu -1$. Thus
$
Var(f)=\omega k/(\nu -1)=k,
$
i.e. the variance does not
depend upon $\nu $ and $\omega .$ When $f\sim GP(h,k)$ $Var(f)=k.$ Under this setting, the first two moments of GP and ETP have the same
form allowing a common interpretation of parameters for the variance for both GPR and eTPR models.
 \cite{r22} also noted that $\nu $ cannot be estimated with a single
realization. In
multivariate t-distribution,  \cite{r6} proposed to use $%
\nu =2$. \cite{r22} suggested that $\nu $ can be chosen according to
investigator's knowledge of robustness of regression error distribution. As $%
\nu \rightarrow \infty $, ETP tends to GP. When robustness property is an
important issue, a smaller $\nu $ is preferred. We tried various values for $%
v$ and find that $v=1.05$ works well. From now on, when $m=1$ we set $v=1.05$
and $\omega =\nu -1=0.05.$

When $m>1$, the eTPR model (\ref{assum1}) includes $m$ random effects $r_{i}
$, $i=1,...,m$. Thus, $\nu $ can be estimted. We have a score equation for $%
\nu $ as follows,
\begin{align}
& \sum_{i=1}^{m}\frac{\partial \log p{_{\beta }(%
\mbox{{\boldmath
${y}$}$_{i}$}|\mbox{{\boldmath ${X}$}$_{i}$})}}{\partial \nu }=-\frac{1}{2}%
\sum_{i=1}^{m}\Big\{\frac{n}{\nu -1}+2\log \Big(1+\frac{\mbox{{\boldmath
${y}$}$_{i}^{T}$}\mbox{{\boldmath
${\Sigma}$}$_{in}^{-1}$}\mbox{{\boldmath
${y}$}$_{i}$}}{2(\nu -1)}\Big)  \notag \\
& -\frac{(n+2\nu )\mbox{{\boldmath ${y}$}$_{i}^{T}$}%
\mbox{{\boldmath
${\Sigma}$}$_{in}^{-1}$}\mbox{{\boldmath
${y}$}$_{i}$}}{2(\nu -1)^{2}+(\nu -1)\mbox{{\boldmath ${y}$}$_{i}^{T}$}%
\mbox{{\boldmath
${\Sigma}$}$_{in}^{-1}$}\mbox{{\boldmath
${y}$}$_{i}$}}-2\psi (\frac{n}{2}+\nu )+2\psi (\nu )\Big\}=0,  \notag
\end{align}
where $\psi (\cdot )$ is a digamma function satisfying $\psi (t+1)=\psi (t)+1/t
$ for $t\in R$.

We assume $h_i(\mbox{\boldmath ${u}$})=0$ as in GPR models. It is straightforward to extend the proposed method to  mean function with general form.

\subsection{Predictive distribution}

Since
\begin{equation}
\left(
\begin{array}{c}
f_{i}(\mbox{{\boldmath ${X}$}$_{i}$}) \\
\mbox{{\boldmath ${y}$}$_{i}$}
\end{array}
\right) \Bigg|\mbox{{\boldmath ${X}$}$_{i}$}\sim EMTD\left( \nu ,\nu
-1,0,\left(
\begin{array}{cc}
\mbox{{\boldmath ${K}$}$_{in}$} & \mbox{{\boldmath ${K}$}$_{in}$} \\
\mbox{{\boldmath ${K}$}$_{in}$} & \mbox{{\boldmath ${\Sigma}$}$_{in}$}
\end{array}
\right) \right) ,  \notag
\end{equation}
from Lemma 2(iii) in Appendix A we have $f_{i}(\mbox{{\boldmath
${X}$}$_{i}$})|\mbox{{\boldmath
${\cal D}$}$_{n}$}\thicksim EMTD(n/2+\nu ,n/2+\nu -1,%
\mbox
{{\boldmath${\mu}$}$_{in}$},\mbox{{\boldmath${C}$}$_{in}$}),$ with
\begin{align}
& \mbox{{\boldmath${\mu}$}$_{in}$}=E(f_{i}(\mbox{{\boldmath ${X}$}$_{i}$})|%
\mbox{{\boldmath ${\cal D}$}$_{n}$})=\mbox{{\boldmath${K}$}$_{in}$}%
\mbox{{\boldmath ${\Sigma}$}$_{in}^{-1}$}\mbox{{\boldmath${y}$}$_{i}$},
\notag \\
& \mbox{{\boldmath${C}$}$_{in}$}=Cov(f_{i}(\mbox{{\boldmath
${X}$}$_{i}$})|\mbox{{\boldmath ${\cal D}$}$_{n}$})=s_{0i}\phi %
\mbox{{\boldmath${K}$}$_{in}$}\mbox{{\boldmath ${\Sigma}$}$_{in}^{-1}$}%
,  \notag \\
& s_{0i}=E(r_{i}|\mbox{{\boldmath ${\cal D}$}$_{n}$})=\frac{%
\mbox{{\boldmath
${y}$}$_{i}^{T}$}\mbox{{\boldmath
${\Sigma}$}$_{in}^{-1}$}\mbox{{\boldmath
${y}$}$_{i}$}+2(\nu -1)}{n+2(\nu -1)}.  \notag
\end{align}
Thus, given $\mbox{\boldmath ${\beta }$},$ $\hat{\mbox{\boldmath ${f}$}}%
_{in}=E(f_{i}(\mbox{{\boldmath
${X}$}$_{i}$})|\mbox{{\boldmath ${\cal D}$}$_{n}$})$ is linear in $%
\mbox{{\boldmath${y}$}$_{i}$},$ i.e. the BLUP for $f_{i}(%
\mbox{{\boldmath
${X}$}$_{i}$}),$ which is an extension of the BLUP in linear mixed models to
eTPR models. This BLUP has a form independent of $\nu$, so that it is also the
BLUP under GPR models. However, the conditional variance depends upon $\nu$. 

For a given new data point $\mbox{\boldmath ${u}$}$, we have
\begin{equation*}
\left(
\begin{array}{c}
\mbox{{\boldmath${y}$}$_{i}$} \\
f_i(\mbox{\boldmath ${u}$})
\end{array}
\right)\Bigg|\mbox{{\boldmath ${X}$}$_{i}$} \sim EMTD\left( \nu ,\nu
-1,0,\left(
\begin{array}{cc}
\mbox{{\boldmath ${\Sigma}$}$_{in}$} &
\mbox{{\boldmath ${k}$}$_{i
u}$} \\
\mbox{{\boldmath ${k}$}$_{i u}^{T}$} & k_i(\mbox{\boldmath ${u}$},%
\mbox{\boldmath ${u}$})
\end{array}
\right) \right),
\end{equation*}
where $\mbox{{\boldmath ${k}$}$_{i u}$}=(k_i(\mbox{{\boldmath ${x}$}$_{i1}$},%
\mbox{\boldmath ${u}$}),\cdots,k_i(\mbox{{\boldmath ${x}$}$_{in}$},%
\mbox{\boldmath ${u}$}))^T$. By Lemma 2(iii), the predictive distribution $p(f_i( \mbox{\boldmath
${u}$})|\mbox{{\boldmath
${\mathcal{D}}$}$_{n}$})$ is $EMTD(n/2+\nu ,n/2+\nu -1,\mu _{in}^{\ast
},\sigma _{in}^{\ast }),$ where
\begin{align}
& \mu _{in}^{\ast }=E(f_i(\mbox{\boldmath ${u}$})|%
\mbox{{\boldmath
${\mathcal{D}}$}$_{n}$})=\mbox{{\boldmath ${k}$}$_{i u}^{T}$}%
\mbox{{\boldmath
${{\Sigma}}$}$_{in}^{{-1}}$}\mbox{{\boldmath${ y}$}$_{i}$},
\label{spec.mean} \\
& \sigma _{in}^{\ast }=Var(f_i(\mbox{\boldmath ${u}$})|%
\mbox{{\boldmath
${\mathcal{D}}$}$_{n}$})=s_{0i}\Big(k_i(\mbox{\boldmath ${u}$},%
\mbox{\boldmath ${u
}$})-\mbox{{\boldmath ${k}$}$_{i u}^{T}$}\mbox{{\boldmath${{%
\Sigma}}$}$_{in}^{-1}$}\mbox{{\boldmath ${k}$}$_{i u}$}\Big).
\label{spec.cov}
\end{align}
Furthermore, from Proposition 1(iii), ${\ f_i|%
\mbox{{\boldmath ${
\mathcal{D}}$}$_{n}$}}\sim ETP{(n/2+\nu ,n/2+\nu -1,\ h_i^{\ast },k_i^{\ast
}),\ } $ {where $h_i^{\ast }(\mbox{\boldmath ${u}$})=\mu^{*}_{in}$ and $%
k_i^{\ast }(\mbox{\boldmath${u}$},\mbox{\boldmath${v}$})=s_{0i}\Big(k_i(%
\mbox
{\boldmath${u}$},\mbox{\boldmath${v}$})-k_{iu}^{T}\mbox{{\boldmath${{%
\Sigma}}$}$_{in}^{-1}$}k_{iv}\Big). $ } From Lemma 2(iii), we also have $y_i(%
\mbox{\boldmath ${u}$})|\mbox{{\boldmath${\mathcal{D}}$}$_{n}$}\thicksim
EMTD(n/2+\nu ,n/2+\nu -1,\mu _{in}^{\ast },\sigma _{in}^{\ast }+s_{0i}\phi )
$ with $E(y_i(\mbox{\boldmath ${u}$})|\mbox{{\boldmath${\mathcal{D}}$}$_{n}$}%
)=\mu _{in}^{\ast }$ and $Var(y_i(\mbox{\boldmath ${u}$})|%
\mbox{{\boldmath${\mathcal{D}}$}$_{n}$})=\sigma _{in}^{\ast }+s_{0i}\phi $.
Consequently, this conditional predictive process can be used to construct
prediction {$\hat{y}_i(\mbox{\boldmath ${u}$})=E(y_i(\mbox{\boldmath ${u}$})|%
\mbox{{\boldmath
${\mathcal{D}}$}$_{n}$})=$}$E(f_i(\mbox{\boldmath ${u}$})|%
\mbox{{\boldmath
${\mathcal{D}}$}$_{n}$})${{\ of the unobserved response $y_i(%
\mbox{\boldmath
${u}$})$ at }}$\mbox{\boldmath ${x}$}=\mbox{\boldmath
${u}$}${\ and its standard error can be formed using the predictive
variance, given by} $\sigma _{in}^{\ast }+s_{0i}\phi $.  The proof is given in
Appendix B.2. The predictive variance for $\hat{f}_i(%
\mbox{\boldmath ${u}$})=\mu_{in}^*$ in (\ref{spec.cov}) differs from that
for $\hat{y}_i(\mbox{\boldmath ${u}$}).$

\cite{r15} discussed the case of $m=1$ and stated that ``both the
predictive mean and the predictive covariance of a TP will differ from that
of a GP after learning kernel hyperparameters''. Their statement is only
partly true when $m=1$. However we show that predictive means and variances  from eTPR and GPR
models always differ  when $m>1$. We consider five commonly used covariance functions here
\citep{r12,r16}: 
squared exponential kernel ($k_{se}$), non-stationary linear kernel ($k_{lin}
$), von Mises-inspired kernel ($k_{vm}$), rational quadratic kernel ($k_{rq}$%
) and Mat$\acute{e}$rn kernel ($k_{m}$) (see the details in Appendix B.2).

\vskip10pt \noindent \textbf{Proposition 2} (\textit{The model with $m=1$})
\begin{itemize}
\item[(i)]  \textit{ Under the kernel functions $k_{se}$, $k_{lin}$ and $%
k_{vm}$, the predictions of $f_{1}(\mbox{{\boldmath ${X}$}$_{1}$})$ and $f_{1}(%
\mbox{\boldmath${u}$})$ from eTPR models are exactly the same as those from GPR
models.}
\item[(ii)] \textit{ Under $k_{rq}$ and $k_{m}$, eTPR and GPR models produce different
predictions. And the predictive
standard error from the eTPR model increases if the
model does not fit the responses well while
that under the GPR model does not depend upon the model fit.}
\end{itemize}

\vskip10pt \noindent \textbf{Proposition 3}   (\textit{The model with $m>1$})\\
\textit{When $m>1$, $\nu$ can be estimated. The eTPR and GPR models under all the five kernel functions discussed above
 have different values of predictions and predictive variances.  The predictive variance under the eTPR model decreases if the model
fits the responses $\mbox{{\boldmath ${y}$}$_{i}$}$ better while that under
the GPR model is still independent of the model fit.}

The proofs of Propositions 2 and 3 are given in Appendix B.2. This paper takes two combinations of $k_{se}+k_{lin}$ and $%
k_{se}+k_{m}$ in simulation studies. For
the first combination $k_{se}+k_{lin}$, eTPR and GPR models with $m=1$ have the exactly same
predictions, but the eTPR has a slightly larger values of predictive variance  than the GPR. For the second case, both predictions and predictive variances are different.


%

Random-effect models consist with three objects, namely the data $%
\mbox{{\boldmath
${\mathcal{D}}$}$_{n}$}$, unobservables (random effects) and parameters
(fixed unknowns) $\mbox{\boldmath ${\beta}$}.$\ For inferences of such
models, \cite{r8} proposed the use of the h-likelihood. \cite{r7} showed that inferences about unobservables allow both Bayesian
and frequentist interpretations. In this paper, we see that the eTPR model
is an extension of random-effect models. Thus, we may view the functional
regression model (\ref{assumed}) either as a Bayesian model, where a GP or
an ETP as a prior, or as a frequentist model where a latent process such as
GP and ETP is used to fit unknown function $f_{01}$ in a functional space
(Chapter 9, \cite{r10}). With the predictive distribution
above, we may form both Bayesian credible and frequentist confidence
intervals. Estimation procedures in Section 3.1\ can be viewed as an
empirical Bayesian method with a uniform prior on $%
\mbox{\boldmath
${\beta}$}.$ In frequentist (or Bayesian) approach, (\ref{margin}) is a
marginal likelihood for fixed (or hyper) parameters.

\section{Robustness and information consistency}

\subsection{Robust properties}

The eTPR models give robust estimates of parameters and the unknown regression function compared to
the GPR models. Even with $m=1$, under some kernel functions
the eTPR models are more robust against outliers in the output space than
the GPR models.

Let $\hat{f}_{iT}(\mbox{\boldmath ${u}$})=\hat{\mu}_{in}^{\ast}=%
\mu_{in}^{*}|_{\mbox{\boldmath
${\beta}$}=\hat{\mbox{\boldmath ${\beta}$}}}$ and $V_{iT}=\hat{{\sigma}}%
_{in}^{\ast }={\sigma}_{in}^{*}|_{\mbox{\boldmath ${\beta}$}=\hat{%
\mbox{\boldmath ${\beta}$}}}$ be the predictive mean and variance for $f_i(\mbox{\boldmath ${u}$})$, under the eTPR model. And let $\hat{%
f}_{iG}(\mbox{\boldmath ${u}$})$ and $V_{iG} $ be those under the GPR model
with $s_{0i}=1$. Let $M_{iT}=(\hat{f}_{iT}(\mbox{\boldmath ${u}$})-f_{i0}(%
\mbox{\boldmath ${u}$}))/\sqrt{V_{iT}}$ and $M_{iG}=(\hat{f}_{iG}(%
\mbox{\boldmath ${u}$})-f_{0i}(\mbox{\boldmath ${u}$}))/\sqrt{V_{iG}}$ be
two student t-type statistics for a null hypothesis $f_i(\mbox{\boldmath
${u}$})=f_{i0}(\mbox{\boldmath ${u}$})$. Under a bounded kernel function, if
$y_{ij}\rightarrow \infty $ for some $j$, $M_{iG}\rightarrow \infty $, while
$M_{iT}$ remains bounded. Therefore, $M_{iT}$ for eTPR is more robust
against outliers in output space compared to that for GPR. This property
still holds for ML estimators.

\vskip10pt \noindent \textbf{Proposition 4} \textit{If kernel functions $k_i$%
, $i=1,\cdots,m$, are bounded, continuous and differentiable on $%
\mbox{{\boldmath ${\theta}$}$_{i}$}$, then for given $\nu$, the ML estimator
$\hat{\mbox{\boldmath
${\beta}$}}$ from the eTPR has bound influence function, while that from the
GPR does not. }

\subsection{Information Consistency}

Let $p_{\phi _{0}}(\mbox{{\boldmath ${y}$}$_{i}$}|f_{0i},%
\mbox{{\boldmath
${X}$}$_{i}$})$ be the density function to generate the data $%
\mbox{{\boldmath ${y}$}$_{i}$}$ given $\mbox{{\boldmath
${X}$}$_{i}$}$ under the true model (\ref{true}), where $f_{0i}$ is the true
underlying function of $f_i$. Let $p_{\theta_i }(f)$ be a measure of random
process $f$ on space ${\mathcal{F}}=\{f(\cdot):\mathcal{X}\rightarrow R\}$.\
Let
\begin{equation*}
p_{\phi ,\theta_i }(\mbox{{\boldmath ${y}$}$_{i}$}|%
\mbox{{\boldmath
${X}$}$_{i}$})=\int_{{\mathcal{F}}}p_{\phi }(\mbox{{\boldmath
${y}$}$_{i}$}|f,\mbox{{\boldmath
${X}$}$_{i}$})dp_{\theta_i }(f),
\end{equation*}
be the density function to generate the data $\mbox{{\boldmath ${y}$}$_{i}$}$
given $\mbox{{\boldmath
${X}$}$_{i}$}$ under the assumed eTPR model (\ref{assum1}). Thus, the
assumed model (\ref{assum1}) is not the same as the true underlying model (%
\ref{true}). Here $\phi $ is the common in both models and $\phi_0$ is the
true value of $\phi$. Let $p_{\phi _{0},\hat \theta_i}(%
\mbox{{\boldmath
${y}$}$_{i}$}|\mbox{{\boldmath ${X}$}$_{i}$})$ be the estimated density
function under the eTPR model. Denote $D[p_{1},p_{2}]=\int (\log {p_{1}}%
-\log {p_{2}})dp_{1}$ by the Kullback-Leibler distance between two densities
$p_{1} $ and $p_{2}$. Then, for fixed $m$, we have the following proposition.

\vskip10pt \noindent \textbf{Proposition 5} \textit{Under the appropriate
conditions in Lemma 3 and  condition (A) of Appendix C, we have for $i=1,\cdots,m$,
\begin{equation*}
\frac{1}{n} E_{\mbox{{\boldmath ${X}$}$_{i}$}}(D[p_{\phi _{0}}(%
\mbox{{\boldmath
${y}$}$_{i}$}|f_{0i},\mbox{{\boldmath ${X}$}$_{i}$}),p_{\phi _{0},\hat{\theta%
}_i}(\mbox{{\boldmath ${y}$}$_{i}$}|\mbox{{\boldmath ${X}$}$_{i}$}%
)])\longrightarrow 0,{\mbox{as}}~~n\rightarrow \infty ,
\end{equation*}
where the expectation is taken over the distribution of $%
\mbox{{\boldmath
${X}$}$_{i}$}$. }

\vskip 10pt

From Proposition 5, the Kullback-Leibler distance between two density
functions for $\mbox{{\boldmath ${y}$}$_{i}$}|\mbox{{\boldmath
${X}$}$_{i}$}$ from the true and the assumed models becomes zero,
asymptotically. Let $\mbox{{\boldmath ${y}$}$_{il}$}=(y_{i1},...,y_{il})^{T}$
and $\mbox{{\boldmath ${X}$}$_{il}$}=(\mbox{{\boldmath ${x}$}$_{i1}$},...,%
\mbox{{\boldmath ${x}$}$_{il}$})^{T}$, $l=1,...,n$. In Appendix C, we show that
\begin{equation}  \label{bspred}
p_{\phi _{0},\theta_i }(\mbox{{\boldmath ${y}$}$_{i}$}|%
\mbox{{\boldmath
${X}$}$_{i}$})=\prod_{l=1}^{n}p_{\phi _{0},\theta_i }(y_{il}|%
\mbox{{\boldmath
${X}$}$_{il}$},\mbox{{\boldmath ${y}$}$_{i(l-1)}$}),
\end{equation}
where
\begin{align}
& p_{\phi _{0},\theta_i }(y_{il}|\mbox{{\boldmath
${X}$}$_{il}$},\mbox{{\boldmath ${y}$}$_{i(l-1)}$})=\int_{{\mathcal{F}}%
}p_{\phi _{0}}(y_{il}|f,\mbox{{\boldmath ${X}$}$_{il}$},\mbox{{\boldmath
${y}$}$_{i(l-1)}$})dp_{\theta_i }(f|\mbox{{\boldmath ${X}$}$_{il}$},%
\mbox{{\boldmath
${y}$}$_{i(l-1)}$}),  \notag \\
& p_{\theta_i }(f|\mbox{{\boldmath ${X}$}$_{il}$},%
\mbox{{\boldmath
${y}$}$_{i(l-1)}$})=\frac{p_{\phi _{0}}(\mbox{{\boldmath ${y}$}$_{i(l-1)}$}%
|f,\mbox{{\boldmath
${X}$}$_{i(l-1)}$})p_{\theta_i}(f)}{\int_{{\mathcal{F}}}p_{\phi _{0}}(%
\mbox{{\boldmath
${y}$}$_{i(l-1)}$}|f^{\prime },\mbox{{\boldmath ${X}$}$_{i(l-1)}$}%
)dp_{\theta_i }(f^{\prime })}.  \notag
\end{align}
Under the true model (\ref{true}), similarly to (\ref{bspred}), we have
\begin{equation*}
p_{\phi _{0}}(\mbox{{\boldmath ${y}$}$_{i}$}|f_{0i},%
\mbox{{\boldmath
${X}$}$_{i}$})=\prod_{l=1}^{n}p_{\phi _{0}}(y_{il}|f_{0i},%
\mbox{{\boldmath
${X}$}$_{il}$},\mbox{{\boldmath ${y}$}$_{i(l-1)}$}).
\end{equation*}
Seeger \textit{et al}. (2008) called $p_{\phi _{0}}(y_{il}|f_{0i},%
\mbox{{\boldmath
${X}$}$_{il}$},\mbox{{\boldmath ${y}$}$_{i(l-1)}$})$ and $p_{\phi _{0},\hat{%
\theta}_i}(y_{il}|\mbox{{\boldmath
${X}$}$_{il}$},\mbox{{\boldmath ${y}$}$_{i(l-1)}$})$ Bayesian prediction
strategies. We show that
\begin{equation*}
D[p_{\phi _{0}}(\mbox{{\boldmath
${y}$}$_{i}$}|f_{0i},\mbox{{\boldmath ${X}$}$_{i}$}),p_{\phi _{0},\hat{\theta%
}_i}(\mbox{{\boldmath ${y}$}$_{i}$}|\mbox{{\boldmath ${X}$}$_{i}$})]=\int
\sum_{l=1}^{n}Q(y_{il}|\mbox{{\boldmath
${X}$}$_{il}$},\mbox{{\boldmath ${y}$}$_{i(l-1)}$})p_{\phi _{0}}(%
\mbox{{\boldmath ${y}$}$_{i}$}|f_{0i},\mbox{{\boldmath ${X}$}$_{i}$})d%
\mbox{{\boldmath ${y}$}$_{i}$},
\end{equation*}
where $Q(y_{il}|\mbox{{\boldmath
${X}$}$_{il}$},\mbox{{\boldmath ${y}$}$_{i(l-1)}$})=\log \{p_{\phi
_{0}}(y_{il}|f_{0i},\mbox{{\boldmath
${X}$}$_{il}$},\mbox{{\boldmath ${y}$}$_{i(l-1)}$})/p_{\phi _{0},\hat{\theta}%
_i}(y_{il}|\mbox{{\boldmath
${X}$}$_{il}$},\mbox{{\boldmath ${y}$}$_{i(l-1)}$})\}$ is a loss function
and $\sum_{l=1}^{n}Q(y_{il}|\mbox{{\boldmath
${X}$}$_{il}$},\mbox{{\boldmath ${y}$}$_{i(l-1)}$})$ is called cumulative
loss. Under the GPR model, Seeger \textit{et al}. (2008)\ and Wang and Shi
(2014) proved information consistency, interpreted it as the average of cumulative
loss $\sum_{l=1}^{n}Q(y_{il}|\mbox{{\boldmath
${X}$}$_{il}$},\mbox{{\boldmath ${y}$}$_{i(l-1)}$})/n$ tending to zero
asymptotically. In this paper, we
show this property for the robust BLUPs. Consequently, the frequentist BLUP
procedure is consistent with the Bayesian strategy in terms of average risk
over an ETP prior.

\section{Numerical studies}

\subsection{Simulation studies}

We use simulation studies to evaluate performance in terms of the robustness for the eTPR model (\ref{assum1}). For GPR and eTPR models, we
use

\begin{itemize}
\item  GPR: $f_{i}\sim GP(0,k_{i})$ and $\epsilon _{ij}\sim N(0,\phi )$, $%
i=1,...,m$, $j=1,...,n$,

\item  eTPR: $f_{i}\sim ETP(\nu ,\nu -1,0,k_{i})$ and $\epsilon _{i}\sim
ETP(\nu ,\nu -1,0,k_{\epsilon})$, $i=1,...,m$,
\end{itemize}

\noindent where $k_i$ are kernel functions, $k_{\epsilon}(%
\mbox{\boldmath ${u}$},\mbox{\boldmath ${v}$})=\phi I(\mbox{\boldmath ${u}$}=%
\mbox{\boldmath ${v}$})$ and $I(\cdot )$ is an indicator function.

Results are based on 500 replications. As we discussed in Section 3.2, when $m=1$,  the eTPR and
GPR methods give the same predictions under
covariance functions such as $k_{se}$, $k_{lin}$ and $k_{vm}$, of $f_1$, but have different prediction values under other kernels such as $k_{rq}$
and $k_{m}$. But the predictions are always different for the two models when $m>1$. Thus, we separate the discussion for  $m=1$ and $m=2$.

\vskip 10pt \noindent \textbf{(i) Models with $m=1$} \vskip 10pt

We used the GPR and eTPR models with kernel function $%
k_1=k_{se}+k_{m}$ to fit the simulation data. Based on the discussion given in the previous section, these two methods with this type of kernel function will result in  different predictions
of $f_1$. When some sparse data points are far away from the dense data points,
predictions at the area of the sparse ones from the eTPR method are
regularized more than those from the LOESS and the GPR methods; i.e. eTPR is a robust method in the input space.

\begin{table}[!ht]
\caption{Mean squared errors of prediction results and their standard
deviation (in parentheses) by the LOESS, GPR and eTPR methods with $m=1$, $%
\protect\phi=0.1$ and $\mbox{\boldmath ${\theta}$}=(0.05,~10,~0.05)$. }
\label{tab1}\tabcolsep=5pt \fontsize{10}{16}\selectfont
 \vskip 0pt
\par
\begin{center}
\begin{tabular}{ccccc} \hline
$n$ & $\sigma^2$ & LOESS & GPR & eTPR \\ \hline
10 & 1 & 0.122(0.121) & 0.116(0.134) & 0.074(0.077) \\
& 2 & 0.198(0.222) & 0.201(0.247) & 0.117(0.140) \\
& 3 & 0.275(0.325) & 0.285(0.345) & 0.162(0.205) \\
& 4 & 0.352(0.428) & 0.360(0.423) & 0.209(0.270) \\ \cline{2-5}
20 & 1 & 0.128(0.154) & 0.120(0.156) & 0.088(0.111) \\
& 2 & 0.228(0.289) & 0.226(0.310) & 0.162(0.212) \\
& 3 & 0.329(0.425) & 0.335(0.462) & 0.237(0.312) \\
& 4 & 0.431(0.562) & 0.443(0.584) & 0.311(0.408) \\ \cline{2-5}
30 & 1 & 0.154(0.196) & 0.139(0.198) & 0.107(0.150) \\
& 2 & 0.284(0.371) & 0.279(0.407) & 0.210(0.294) \\
& 3 & 0.414(0.547) & 0.421(0.604) & 0.312(0.430) \\
& 4 & 0.544(0.723) & 0.568(0.787) & 0.414(0.563) \\ \hline
\end{tabular}
\end{center}
\end{table}

\begin{figure}[h!]
\begin{center}
\includegraphics[height=0.7\textwidth,width=0.98\textwidth]{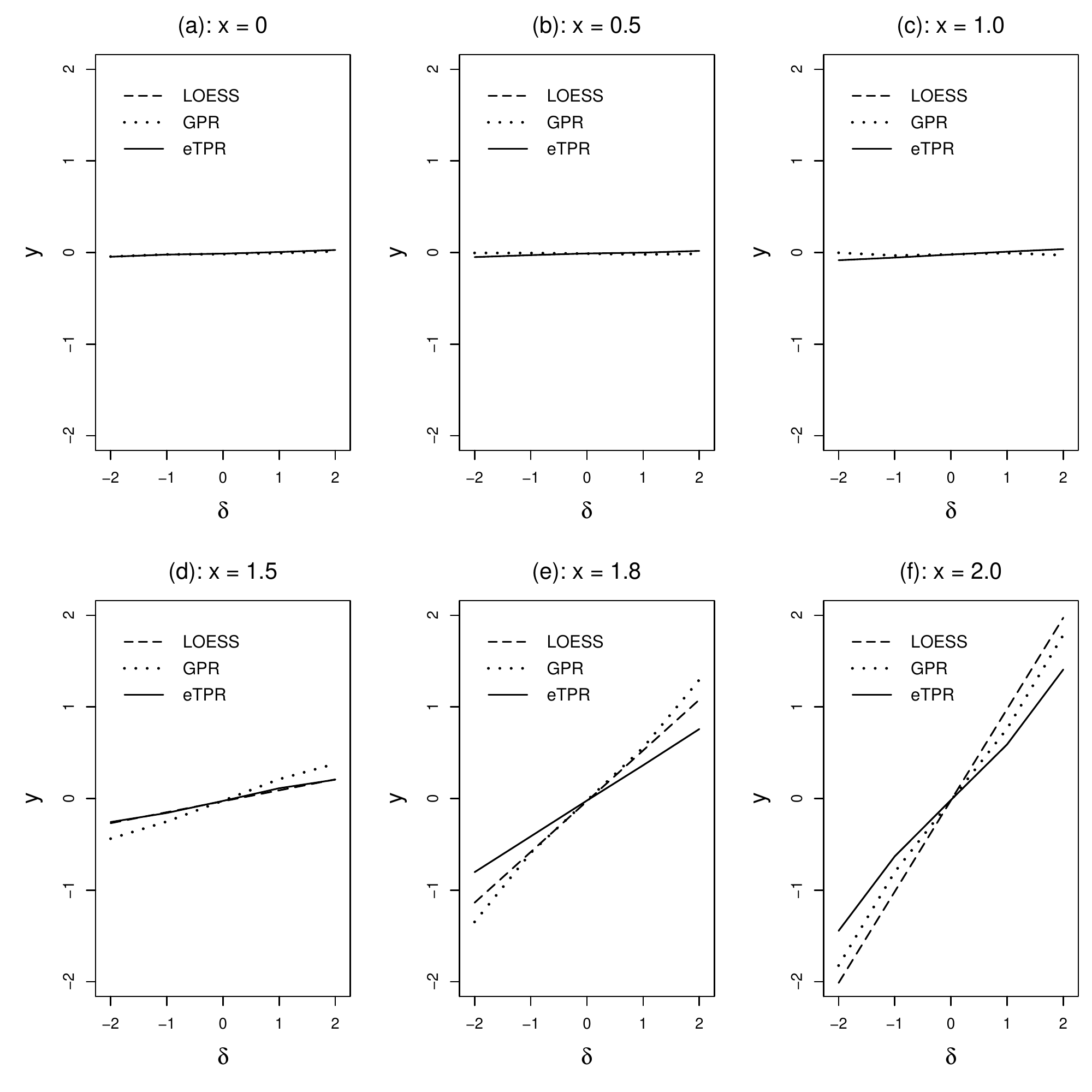}
\end{center}
\par
\caption{Predicted values at the data points $x\in%
\{0.0,~0.5,~1.0,~1.5,~1.8,~2.0\} $ with $m=1$ and constant disturbance at
the point 2.0, where mean function is 0, and dashed, doted and solid lines
respectively represent predictions from the LOESS, GPR and eTPR methods. }
\label{fig2}
\end{figure}

We first generate data
from the process regression model (\ref{assumed}).  We assume $f_{1}$ follows a GP with
mean $0$ and the kernel function $k_{se}+k_{lin}$, and error term follows a
normal distribution with mean 0 and variance $\phi $. We set $\phi =0.1$ and $\mbox{\boldmath ${\theta}$}=(\eta _{0},\eta
_{1},\xi _{0})=(0.05,~10,~0.05)$, where $\eta _{0}$ and $\eta _{1}$ are
parameters for the kernel $k_{se}$, and $\xi _{0}$ is the one in the kernel $k_{lin}$. In each replication of the simulation study, $N=61$ points
evenly spaced in [0,~2.0] are generated  and used for covariate, denoted by $S$, where
the $46$th and $61$th points are 1.5 and 2.0. We take $n-1$ points with
orders evenly spaced in the first 46 points of $S$
and point $2.0$ as the training data, and the remaining as the test data, where $n$ is the sample size. Three different sizes, $n=10$, 20 and 30, are considered.
The test data points are denoted by $\{x_{j}^{\ast }:j=1,...,J\}$ with $J=N-n$. To  show the performance of robustness, the training data at point 2.0 is disturbed by adding extra errors generated from $N(0, \sigma^2)$ with $\sigma^2=1, 2, 3$ and $4$ respectively.

We show the results from one replication  with $n=10$ and $\sigma^2=2$ and 4 in Figure \ref{fig1}. The prediction and its 95\% prediction point-wise confidence intervals are computed and presented.
From Figure 1 we see that
the eTPR method has selective shrinkage (Wauthier and Jordan, 2010) and
gives a wider interval in the area with sparse data points.

The simulation study result based on 500 replications are reported in
Table \ref{tab1}. The performance is measured by the  mean
squared error between the predictions and the true values for  the test data:   $MSE=\sum_{j=1}^{J}(\hat{f}_1(x_{j}^{\ast
})-f_{10}(x_{j}^{\ast }))^{2}/J$. Table \ref{tab1} shows that the eTPR model performs the best among the three methods: LOESS,
GPR and eTPR, and the GPR has comparable MSE with the LOESS, in the presence of outliers. The improvement
is greater with large $\sigma ^{2}$ as expected. This is also confirmed in Figure \ref{fig2}.  Instead of random disturbance, a constant disturbance $\delta$ is added to the training
data point 2.0,  where $\delta =-2,-1,0, 1$ and $2$. Predicted values $\hat{y}_1(u)$ at data
points $u=0,0.5,1.0,1.5,1.8$ and $2.0$ are calculated respectively by the LOESS, GPR and
eTPR. Figure \ref{fig2} presents the average value of
predictions based on 500 replications at each data points against $\delta$. It shows clearly that the influence from outliers is ignorable for the data points in the area less than 1.5, in which there are densely observed data. But the influence becomes larger in the sparse data area $1.5<x\leq 2.0$.  Predictions from the eTPR method are shrunken more heavily than those calculated from the other methods in this area.

\begin{table}[!ht]
\caption{Mean squared errors of prediction results and their standard
deviation (in parentheses) by the LOESS, GPR and eTPR methods with $m=1$ and no outliers.}
\label{Dtab2}\tabcolsep=5pt \fontsize{10}{16}\selectfont
 \vskip 0pt
\par
\begin{center}
\begin{tabular}{ccccc} \hline
$n$&Model & LOESS & GPR & eTPR \\ \hline
10&(1)&0.044(0.028)&0.034(0.025)&0.035(0.025)\\
&(2)&0.046(0.029)&0.046(0.029)&0.044(0.029)\\
20&(1)&0.020(0.012)&0.022(0.018)&0.021(0.015)\\
&(2)&0.024(0.014)&0.030(0.018)&0.028(0.017)\\
30&(1)&0.013(0.009)&0.013(0.011)&0.014(0.009)\\
&(2)&0.016(0.010)&0.021(0.013)&0.020(0.012)\\
\hline
\end{tabular}
\end{center}
\end{table}

\begin{table}[!ht]
\caption{Mean squared errors of prediction results and their standard
deviation (in parentheses) by the LOESS, GPR and eTPR methods with 6
different simulation setups and $m=1$.}
\label{tab2}\tabcolsep=5pt \fontsize{10}{16}\selectfont
 \vskip 0pt
\par
\begin{center}
\begin{tabular}{ccccc} \hline
$n$ & Model & LOESS & GPR & eTPR \\ \hline
10 & (1) & 1.241(4.955) & 0.158(0.240) & 0.089(0.135) \\
& (2) & 1.027(3.643) & 0.188(0.307) & 0.146(0.256) \\
& (3) & 0.412(1.584) & 0.131(0.198) & 0.078(0.077) \\
& (4) & 0.415(1.586) & 0.147(0.192) & 0.110(0.119) \\
& (5) & 1.111(4.325) & 0.204(0.456) & 0.122(0.262) \\
& (6) & 1.232(3.553) & 0.365(0.400) & 0.327(0.443) \\ \cline{2-5}
20 & (1) & 0.682(4.497) & 0.074(0.125) & 0.047(0.064) \\
& (2) & 0.515(2.350) & 0.089(0.151) & 0.074(0.117) \\
& (3) & 0.160(0.621) & 0.073(0.088) & 0.058(0.061) \\
& (4) & 0.164(0.621) & 0.086(0.089) & 0.081(0.094) \\
& (5) & 0.290(1.163) & 0.086(0.162) & 0.068(0.153) \\
& (6) & 0.454(1.372) & 0.298(0.520) & 0.270(0.432) \\ \cline{2-5}
30 & (1) & 0.793(6.723) & 0.043(0.082) & 0.033(0.060) \\
& (2) & 0.324(2.643) & 0.060(0.119) & 0.052(0.107) \\
& (3) & 0.186(1.838) & 0.053(0.060) & 0.045(0.043) \\
& (4) & 0.189(1.843) & 0.067(0.070) & 0.062(0.063) \\
& (5) & 0.366(2.764) & 0.063(0.144) & 0.055(0.127) \\
& (6) & 0.533(2.099) & 0.256(0.336) & 0.246(0.314) \\ \hline
\end{tabular}
\end{center}
\end{table}

We also investigate robust property  against model misspecification and/or outliers
in output space. Data $y_{1j}$ are generated from the following 6 process models:\newline
(1) $f_1\sim GP(0,k)$, $\epsilon_1 \sim N(0,\phi )$, $\phi=0.1$, and $%
\mbox{\boldmath ${\theta}$}=(0.05,~2,~0.05)$;\newline
(2) $f_1\sim GP(0,k)$, $\epsilon_1 \sim N(0,\phi )$, $\phi=0.1$, and $%
\mbox{\boldmath ${\theta}$}=(0.1,~4,~0.1)$;\newline
(3) $f_1\sim GP(0,k)$, $\epsilon_1 \sim \sqrt{\phi} t_{2}$, $\phi=0.1$, and $%
\mbox{\boldmath ${\theta}$}=(0.05,~2,~0.05)$;\newline
(4) $f_1\sim GP(0,k)$, $\epsilon_1 \sim \sqrt{\phi} t_{2}$, $\phi=0.1$, and $%
\mbox{\boldmath ${\theta}$}=(0.1,~4,~0.1)$;\newline
(5) $f_1\sim ETP(2,2,0,k)$, $\epsilon_1 \sim ETP(2,2,0,k_{\epsilon})$, $\phi=0.1
$, and $\mbox{\boldmath ${\theta}$}=(0.05,~2,~0.05)$;\newline
(6) $f_1$ and $\epsilon_1$ has a joint ETP (\ref{assum1}) with $\phi=0.1$
and $\mbox{\boldmath ${\theta}$}=(0.05,~2,~0.05)$,\newline
where $k=k_{se}+k_{lin}$. 
We also take sample size as $n=10$, 20 and 30. In each replication, $N=50$ points are
generated evenly spaced in [0,~3.0], denoted by $S$. And $n$ points evenly
spaced in $S$ 
are taken as training data, and the remaining as testing data. Values of mean
squared error (MSE) for test data are computed for each of the LOESS, GPR and eTPR methods. 
{  Table \ref{Dtab2} presents MSEs of LOESS, GPR and eTPR under Cases (1) and (2), where the data are
generated from the GPR models and do not exists outliers. It shows that eTPR has comparable performance with GPR. As expected, LOESS is comparable with or slightly better than the other two models for those two simple cases. } To study robustness, we
consider  model misspecifications and outliers in the following ways.
 For Cases (1), (2) (5) and (6), the $n$th
data point from the training data set is added with a $t_{1}$ error. Thus, Cases (1) and (2) have outliers,
Cases (3) and (4) have non-normal errors and Cases (5) and (6) have both. We
see from Table \ref{tab2} that the eTPR method performs consistently better than the other two methods, especially when sample size is small.
{  More simulation study results are presented in Table D.11 in Appendix D.

Performance of the eTPR is also investigated by studying data generation model with peak contamination (\cite{dr1}).
Under Cases (1), (2), (5) and (6), data $y_{1}(u)$ are contaminated through $\tilde y_{1}(u)=y_{1}(u)+4c_1\eta_1$ if $T_1\leq u\leq T_1+1/15$, otherwise $\tilde y_{1}(u)=y_{1}(u)$,
where $c_1$ is 1 with probability 0.8 and 0 with probability 0.2, $\eta_1$ is independent of $c_1$ taking value of 1 or -1 with
equal probability of 0.5, $T_1$ is random number from a uniform distribution in [0,14/15]; see the details in (\cite{dr1}). Sample size is $n=20$.
MSEs of prediction from the LOESS, GPR and eTPR are presented in Table \ref{Dtab3}. It shows that eTPR has the smallest MSE, and GPR performs better than LOESS.}

\begin{table}[!ht]
\caption{Mean squared errors of prediction results and their standard
deviation (in parentheses) by the LOESS, GPR and eTPR methods under peak contamination and $n=20$.}
\label{Dtab3}\tabcolsep=5pt \fontsize{10}{16}\selectfont
 \vskip 0pt
\par
\begin{center}
\begin{tabular}{ccccc} \hline
&Model & LOESS & GPR & eTPR \\ \hline
&(1)&0.142(0.097)&0.085(0.090)&0.065(0.070)\\
&(2)&0.145(0.099)&0.105(0.087)&0.092(0.083)\\
&(5)&0.168(0.115)&0.119(0.112)&0.101(0.120)\\
&(6)&0.353(0.359)&0.303(0.365)&0.286(0.393)\\

\hline
\end{tabular}
\end{center}
\end{table}

\begin{figure}[h!]
\begin{center}
\includegraphics[height=0.7\textwidth,width=0.98\textwidth]{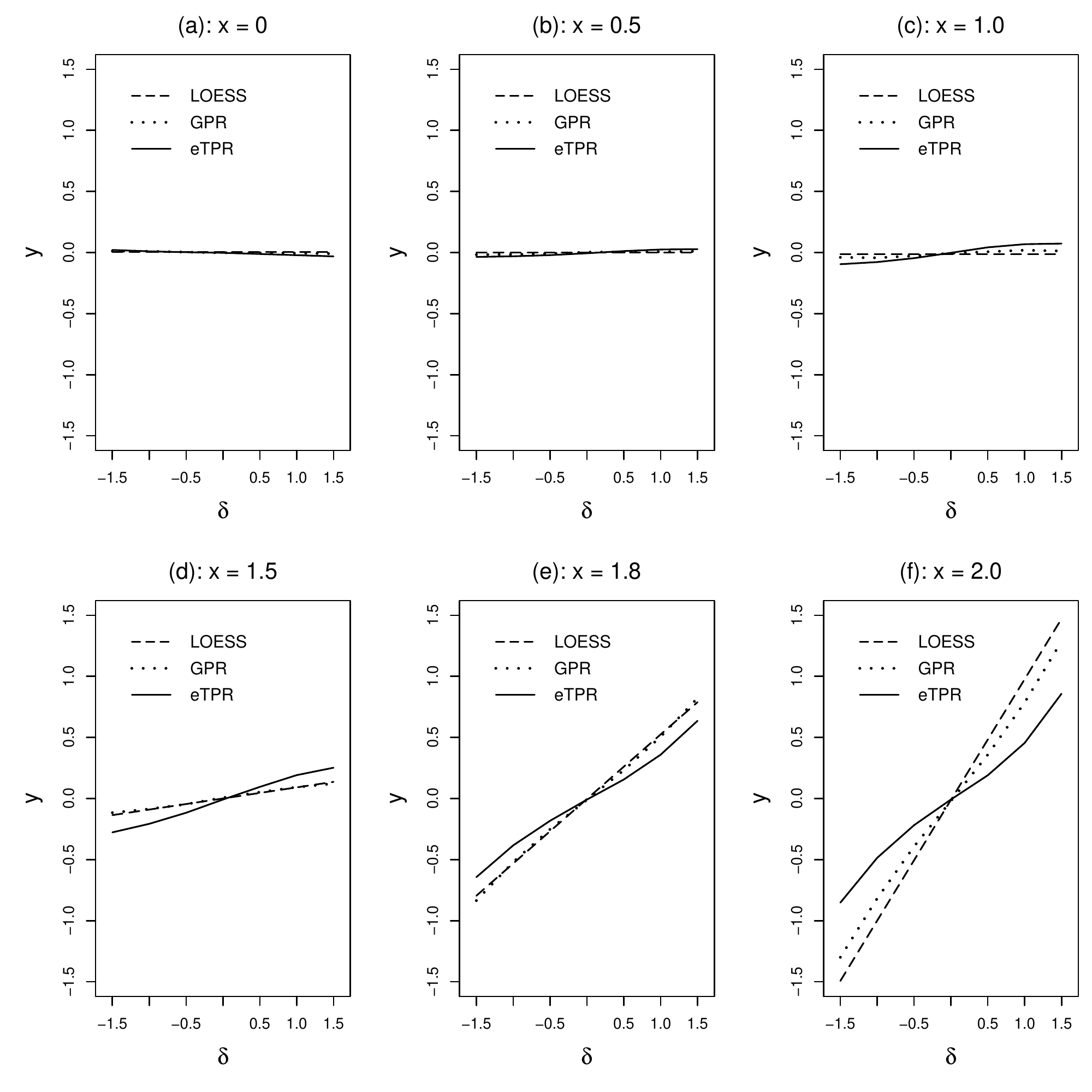}
\end{center}
\par
\caption{Predicted values at the data points $x\in%
\{0.0,~0.5,~1.0,~1.5,~1.8,~2.0\} $ with $m=2$ and constant disturbance at
the point 2.0, where dashed, doted and solid lines respectively represent
predictions from the LOESS, GPR and eTPR methods, respectively. }
\label{fig3}
\end{figure}

\vskip 10pt \noindent \textbf{(ii) Models with $m=2$} \vskip 10pt

Now we study the models when $m>1$. We take
$m=2$ and kernel functions, $k_i=k_{se}+k_{lin}, i=1,2$, for both GPR and
eTPR models. Based on the previous discussion,  GPR and eTPR models under this kernel have different
predictions when $m>1$, while they have the same ones when $m=1$.

Let us investigate selective shrinkage property of the eTPR firstly.
Simulation data are generated similar to  the cases of $%
m=1$, where $f_i$, $\epsilon_i$, $x_{ij}$ and $y_{ij}$ for each $i$ follow
the same setups as those in the constant disturbance case  above.
Sample sizes are $n_1=n_2=n=10$. The average predicted values at data points $u=0,0.5,1.0,1.5,1.8$ and $2.0
$ are presented in Figure \ref{fig3}.  It shows the  results similar to  the case of $%
m=1$ in Figure  \ref{fig2}. The eTPR model shrinks the prediction at sparse data region much more than the
LOESS and GPR models.

\begin{table}[th]
\caption{Mean squared errors of prediction results and their standard
deviation (in parentheses) by the LOESS, GPR and eTPR methods with $m=2$ and
without outliers. Dimension of the covariates is 1 or 3.}
\label{tab3}\tabcolsep=5pt \fontsize{10}{16}\selectfont
 \vskip 0pt
\par
\begin{center}
\begin{tabular}{cccccc} \hline
Dimension & Model & LOESS & GPR & eTPR($\nu=1.05$) & eTPR \\ \hline
1 & (1) & 0.033(0.034)&0.021(0.015)&0.020(0.015)&0.022(0.016) \\
& (2) & 0.065(0.069)&0.043(0.032)&0.042(0.030)&0.041(0.030) \\
\cline{2-6}
3 & (1) & 19.111(155.179)&0.061(0.037)&0.059(0.036)&0.061(0.042) \\
& (2) & 19.518(154.742)&0.098(0.061)&0.096(0.061)&0.103(0.063) \\
\hline
\end{tabular}
\end{center}
\end{table}

\begin{table}[!ht]
\caption{Mean squared errors of prediction results and their standard
deviation (in parentheses) by the LOESS, GPR and eTPR methods with $m=2$ and outliers.
Dimension of the covariates is 1 or 3.}
\label{tab4}\tabcolsep=5pt \fontsize{10}{16}\selectfont
 \vskip 0pt
\par
\begin{center}
\begin{tabular}{cccccc}\hline
Dimension & Model & LOESS & GPR & eTPR($\nu=1.05$) & eTPR \\ \hline
1&(1)&0.257(0.753)&0.160(0.464)&0.152(0.475)&0.139(0.441)\\
&(2)&0.293(0.815)&0.183(0.482)&0.171(0.464)&0.161(0.444)\\
&(3)&0.253(0.799)&0.147(0.511)&0.156(0.760)&0.130(0.495)\\
&(4)&0.398(0.894)&0.247(0.735)&0.227(0.688)&0.222(0.710)\\
&(5)&0.392(1.157)&0.199(0.577)&0.155(0.351)&0.154(0.389)\\
&(6)&0.389(0.555)&0.300(0.435)&0.295(0.440)&0.277(0.440)\\
\cline{2-6}
3&(1)&26.216(253.396)&0.272(0.919)&0.264(0.909)&0.252(0.922)\\
&(2)&68.825(672.400)&0.246(0.458)&0.242(0.456)&0.230(0.434)\\
&(3)&33.026(412.647)&0.329(0.923)&0.318(0.841)&0.308(0.919)\\
&(4)&25.892(177.680)&0.318(0.503)&0.312(0.446)&0.294(0.430)\\
&(5)&29.432(281.501)&0.298(0.613)&0.293(0.610)&0.285(0.624)\\
&(6)&24.803(219.508)&0.499(0.719)&0.486(0.661)&0.475(0.678)\\

\hline
\end{tabular}
\end{center}
\end{table}

\begin{table}[!ht]
\caption{Estimates of $\protect\nu$ with $m=2$ and 1- or 3- dimensional covariates}
\label{tab5}\tabcolsep=3pt \fontsize{8}{12}\selectfont
 \vskip 0pt
\par
\begin{center}
\begin{tabular}{ccccccc} \hline
Dimension & (1) & (2) & (3) & (4) & (5) & (6) \\ \hline
1 & 3.592(0.912) & 3.712(0.803) & 3.557(0.938) & 3.693(0.819)  & 3.553  (0.951)
& 3.471(1.030) \\ \cline{2-7}
3 & 3.782(0.692) &  3.734(0.742) &  3.804(0.650) & 3.736(0.768) & 3.666(0.854)
& 3.666(0.832) \\ \hline
\end{tabular}
\end{center}
\end{table}

We now consider the models with  one
single explanatory variate in function $f_i$, $i=1,2$. Sample sizes are taken as $%
n_1=n_2=n=10$. Similar to $m=1$, data $y_{ij}$ are generated from the
following 6 process models:\newline
(1) $f_i\sim GP(0,k)$, $\epsilon_i \sim N(0,\phi )$, $\phi=0.05$, and $%
\mbox{\boldmath ${\theta}$}=(0.025,~2,~0.025)$;\newline
(2) $f_i\sim GP(0,k)$, $\epsilon_i \sim N(0,\phi )$, $\phi=0.1$, and $%
\mbox{\boldmath ${\theta}$}=(0.05,~2,~0.05)$;\newline
(3) $f_i\sim GP(0,k)$, $\epsilon_i \sim \sqrt{\phi} t_{2}$, $\phi=0.05$, and
$\mbox{\boldmath ${\theta}$}=(0.025,~2,~0.025)$;\newline
(4) $f_i\sim GP(0,k)$, $\epsilon_i \sim \sqrt{\phi} t_{2}$, $\phi=0.1$, and $%
\mbox{\boldmath ${\theta}$}=(0.05,~2,~0.05)$;\newline
(5) $f_i\sim ETP(2,2,0,k)$, $\epsilon_i \sim ETP(2,2,0,k_{\epsilon})$, $%
\phi=0.05$, and $\mbox{\boldmath ${\theta}$}=(0.025,~2,~0.025)$;\newline
(6) $f_i$ and $\epsilon_i$ has a joint ETP (\ref{assum1}) with $\phi=0.05$
and $\mbox{\boldmath ${\theta}$}=(0.025,~2,~0.025)$,\newline
where $k=k_{se}+k_{lin}$, $\mbox{\boldmath ${\theta}$}=(\eta
_{0},\eta_{1},\xi _{0})$. 
As before,  $N=50$ points evenly spaced in [0,~3.0] are generated,  $n=10$ randomly selected points are used  as training data and the remaining as testing
data.  Here the parameter $\nu$ is estimated. As comparison, we also consider the model with fixed value of $\nu=1.05$ (as in the case of $m=1$),  denoted by eTPR($\nu=1.05$). Values of mean squared error based on 500 replications are reported for all the four models in the upper panel  in Table \ref{tab3}. It  shows the results  for Cases (1)
and (2), where the data are generated from GPR models. We can see that all
four methods performs similarly.

To study robustness, in Cases (1), (2), (5) and (6), one data point
for each group is randomly selected from the training data set and is added
with a $t_{2}$ error. We see from the upper panel  in Table \ref{tab4}
that eTPR methods perform better than the LOESS and GPR. And eTPR has smaller MSE than eTPR($\nu=1.05$), indicating a better fit when the parameter $\nu$ can be estimated from the data.

We also study  the models with multivariate
covariate $\mbox{\boldmath ${x}$}=(x_{1},x_{2},x_{3})^{T}$. In this case, $%
\mbox{\boldmath ${\theta}$}=(\eta _{0}, \eta _{1},\eta _{2},\eta _{3},
\xi_{0},\xi_{1},\xi_{2}),$ where $(\eta _{0}, \eta _{1},\eta _{2},\eta _{3})$
are the parameters in the kernel $k_{se}$, and ($\xi_{0},\xi_{1},\xi_{2}$) are
the ones in the $k_{lin}$. To generate data, we follow the previous six
process models, but $\mbox{\boldmath ${\theta}$}%
=(0.05,~2,~2,~2,~0.05,~0.05,~0.05)$ for Cases 1, 3, 5 and 6, and for Cases 2
and 4, $\mbox{\boldmath ${\theta}$}=(0.1,~4, ~4,~4,~0.1,~0.1,~0.1)$. Let $%
S_{1}$, $S_{2}$ and $S_{3}$ be sets of $N=50$ points evenly spaced in the
intervals (-2, 2), (0, 3) and (1, 2), respectively. For each group, we
randomly take $10$ points as the training data and the remaining as the test
data. The lower panel in Table \ref{tab3} shows the MSEs for Cases (1)
and (2), where GPR models are the true models of the generated data. We can see
that the eTPR methods have comparable MSEs with the GPR, both perform pretty well. But the LOESS
method fails in this case this is because it is designed to deal with the model with a one-dimensional covariate only.

To study robustness, we also randomly select one data point for each group from the
training data and add  $t_{2}$ errors for Cases (1), (2), (5)
and (6). The lower panel in Table \ref{tab4} presents the values of MSE. Again, the eTPR methods perform better than the GPR method, and the LOESS fails.

Table \ref{tab5} lists estimates of $\nu$ for 1- and 3- dimensional
covaraite. We can see that the estimates of $\nu$ are much larger than $\nu=1.05$, the fixed value we used in the setting and in the case of $m=1$.

More curves are generated to study the performance of prediction from the 4 methods. We take $m=10$ and simulation cases (1) - (6)
are the same as those in the situation of $m=2$ and 3-dimensional covariates. For Cases (1), (2), (5) and (6), two curves from $m$ ones are selected, and for each selected curve,
one data point is randomly chose from the training data sets and added with a $t_{2}$ error. Table \ref{tab4-1} presents the values of MSE and
Table \ref{tab5-1} lists estimates of $\nu$. These tables shows the similar conclusion with $m=2$.
Moreover, MSEs and their standard deviations for GPR and eTPR methods become smaller compared to $m=2$.

\begin{table}[!ht]
\caption{Mean squared errors of prediction results and their standard
deviation (in parentheses) by the LOESS, GPR and eTPR methods with $m=10$ and 3-dimension covariates.}
\label{tab4-1}\tabcolsep=5pt \fontsize{10}{16}\selectfont
 \vskip 0pt
\par
\begin{center}
\begin{tabular}{ccccc}\hline
 Model & LOESS & GPR & eTPR($\nu=1.05$) & eTPR \\ \hline

(1)&27.167(204.574)&0.131(0.547)&0.126(0.543)&0.124(0.541)\\
(2)&67.839(568.164)&0.164(0.463)&0.161(0.488)&0.154(0.395)\\
(3)&39.760(373.899)&0.335(0.647)&0.333(0.640)&0.323(0.653)\\
(4)&64.973(517.224)&0.383(0.893)&0.336(0.563)&0.328(0.579)\\
(5)&39.024(348.024)&0.146(0.201)&0.142(0.182)&0.136(0.132)\\
(6)&40.297(350.117)&0.376(0.300)&0.373(0.303)&0.368(0.298)\\

\hline
\end{tabular}
\end{center}
\end{table}

\begin{table}[!ht]
\caption{Estimates of $\protect\nu$ with $m=10$ and 3- dimensional covariates}
\label{tab5-1}\tabcolsep=3pt \fontsize{8}{12}\selectfont
 \vskip 0pt
\par
\begin{center}
\begin{tabular}{cccccc} \hline
 (1) & (2) & (3) & (4) & (5) & (6) \\ \hline
3.895(0.393) &3.897(0.417) &3.554(0.820) &3.575(0.801)& 3.480(0.901) & 3.401(0.922)\\ \hline
\end{tabular}
\end{center}
\end{table}

\vskip 10pt

\subsection{Real examples}

\vskip 10pt

\begin{figure}[h!]
\begin{center}
\includegraphics[height=0.7\textwidth,width=0.98\textwidth]{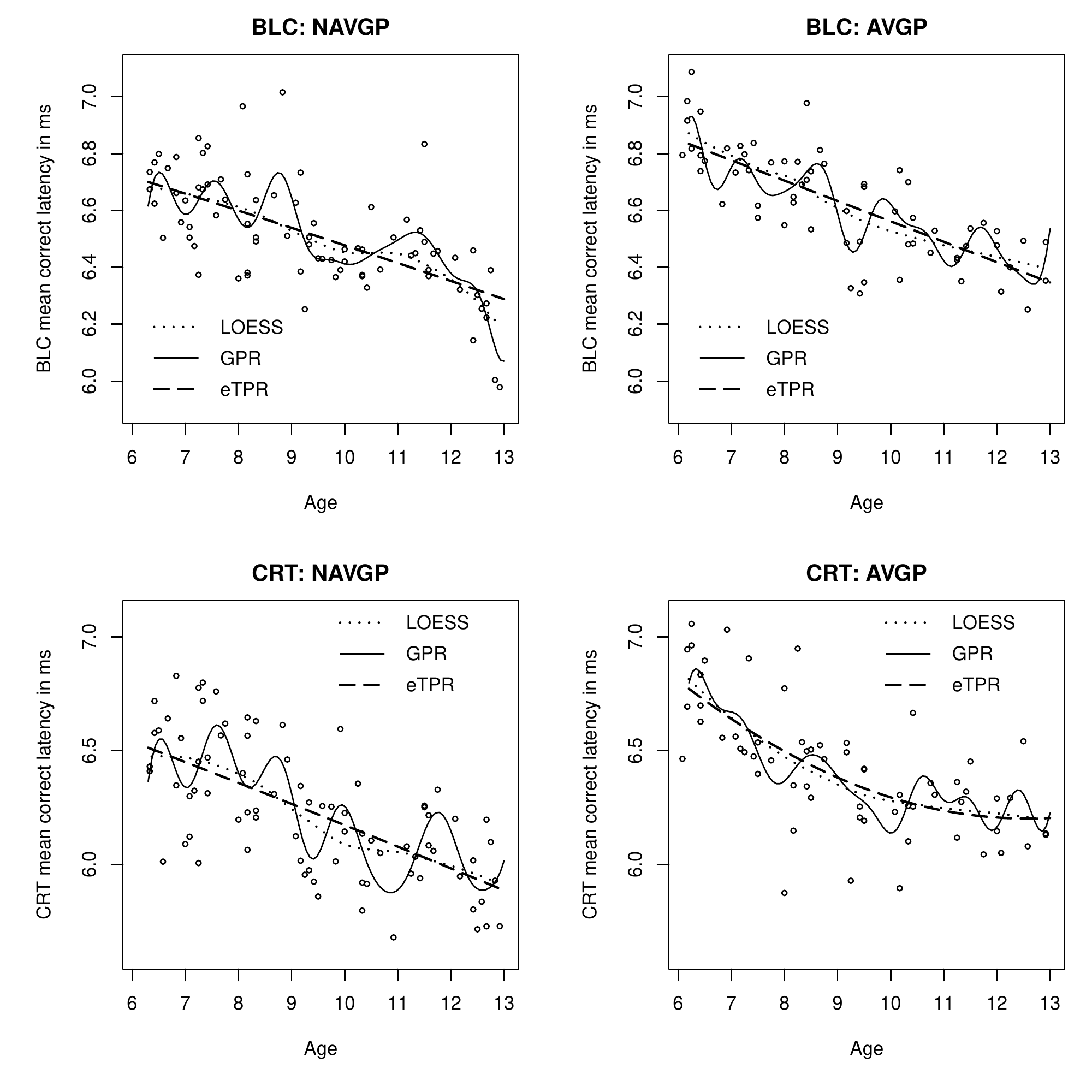}
\end{center}
\caption{Prediction curves from the LOESS, GPR and eTPR methods with kernel $%
k_i=k_{se}+k_{lin}$ for the BLC and CRT data, where circles represent data
points, and solid, dotted and dashed lines stand for predictions from the
GPR, LOESS and eTPR, respectively. }
\label{fig4}
\end{figure}

\begin{figure}[h!]
\begin{center}
\includegraphics[height=0.7\textwidth,width=0.98\textwidth]{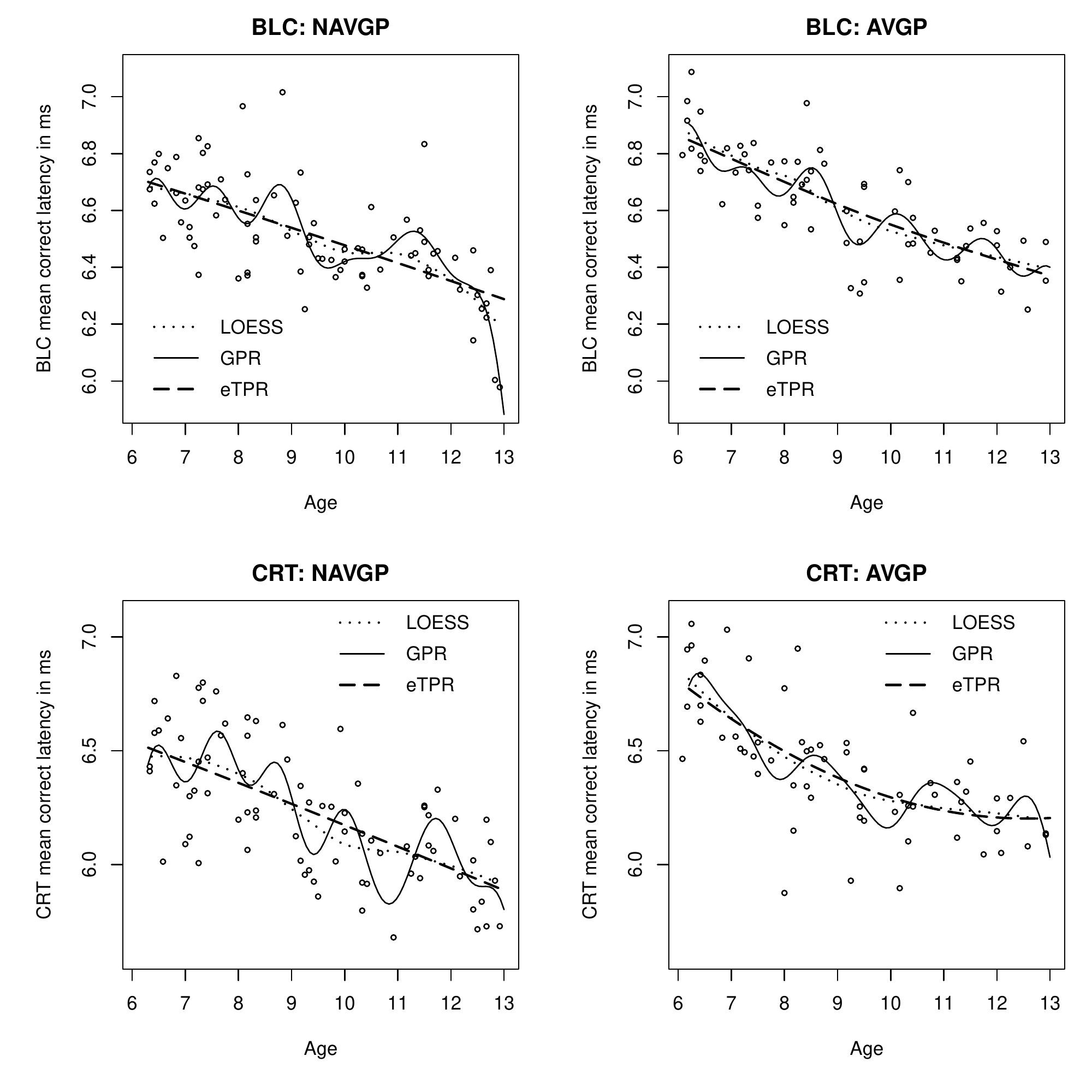}
\end{center}
\caption{Prediction curves from the LOESS, GPR and eTPR methods with kernel $%
k_i=k_{se}+k_{m}$ for the BLC and CRT data, where circles represent data
points, and solid, dotted and dashed lines stand for predictions from the
GPR, LOESS and eTPR, respectively. }
\label{fig5}
\end{figure}


\begin{table}[!ht]
\caption{Prediction errors and their standard deviation (in parentheses) for
the 3 real data sets by the LOESS, GPR and eTPR methods.}
\label{tab6}\tabcolsep=5pt \fontsize{10}{16}\selectfont
 \vskip 0pt
\par
\begin{center}
\begin{tabular}{ccccc} \hline
kernel & Data & LOESS & GPR & eTPR \\ \hline
$k_{se}+k_{lin}$ & BLC & 0.019(0.006) & 0.025(0.017) & 0.020(0.006) \\
& CRT & 0.049(0.012) & 0.062(0.026) & 0.049(0.012) \\
& Snow & 1.142(0.102) & - & 1.111(0.105) \\
& Spatial & 0.510(0.119) & - & 0.193(0.075) \\ \cline{2-5}
$k_{se}+k_{m}$ & BLC & 0.019(0.006) & 0.021(0.009) & 0.020(0.006) \\
& CRT & 0.049(0.012) & 0.054(0.013) & 0.049(0.012) \\
& Snow & 1.142(0.102) & 1.113(0.105) & 0.953(0.164) \\
& Spatial & 0.510(0.119) & 0.203(0.082) & 0.197(0.079) \\ \hline
\end{tabular}
\end{center}
\end{table}

The eTPR model (\ref{assum1}) is applied to executive function research data
coming from the study in children with Hemiplegic Cerebral Palsy and
consisting of 84 girls and 57 boys from primary and secondary schools. These
students were subdivided into two groups ($m=2$): the action video game
players group (AVGPs) (56\%) and the non action video game players group
(NAVGPs) (44\%). To demonstrate the proposed method, we take 2 measurement
indices, Big/Little Circle (BLC) mean correct latency and Choice Reaction
Time (CRT) mean correct latency which are investigated as age of children:
for more details of this data set, see Xu \textit{et al}. (2015). Before
applying the proposed methods, we take logarithm of BLC and CRT mean correct
latencies. For the GPR and eTPR methods, kernel function  $%
k_i=k_{se}+k_{lin}$ or $k_{se}+k_{m}$ is used. Figures \ref{fig4} and \ref{fig5}
present prediction curves under these two kernels, where circles represent
observed data points, and solid line, dashed line and dotted line stand for
predictions from the GPR, eTPR and LOESS methods, respectively. Estimates of
$\nu$ are 6 for eTPR models with either  kernel function. We
can see prediction curves from the LOESS and eTPR methods are more smooth
than those from the GPR method. Furthermore, prediction curves from the GPR
method are more influenced by the choice of kernel function compared to those
from the eTPR.

We randomly select 80\% observation as training data and compute prediction
errors for the remaining data points (i.e. the test data). This procedure is
repeated 500 times. Table \ref{tab6} presents mean prediction errors of BLC
and CRT mean correct latencies. We can see that the GPR is the worst, while
the eTPR is comparable with the LOESS. Again, prediction errors from the GPR
more depend on kernel functions compared to the eTPR. Thus, selection of
kernel function is more important for the GPR method.

We also apply the three methods to Whistler snowfall data and
spatial interpolation data. Those are the models with $m=1$. Whistler snowfall data contain daily snowfall
amounts in Whistler for the years 2010 and 2011, and can be downloaded at
http://www.climate.weather office.ec.gc.ca. Response for snow data is
logarithm of (daily snowfall amount+1) and covariate is time. For spatial
interpolation data, rainfall measurements at 467 locations were recorded in
Switzerland on 8 May 1986, and can be found at http://www.ai-geostats.org
under SIC97. Spatial interpolation data has response, logarithms of
(rainfall amount+1), and two covariates for coordinates of location. We also take
kernel function $k_1=k_{se}+k_{lin}$ or $k_{se}+k_{m}$ for both the GPR and eTPR
methods. Prediction errors of the GPR, eTPR and LOESS are listed in Table
\ref{tab6}. When $m=1$, results of the eTPR are the same as those of the
GPR for kernel $k_{se}+k_{lin}$, so we only list prediction errors of the
eTPR in this table. We can see that eTPR has the smallest prediction error.
For spatial data with multivariate predictors, the LOESS is much
worse. Overall, the eTPR is the best in prediction. 
%

\section{Concluding remarks}

Advantages of a GPR model include that it offers a nonparametric regression
model for data with multi-dimensional covariates, the specification of
covariance kernel enables to accommodate a wide class of nonlinear
regression functions, and it can be applied to analyze many different types
of data including functional data. In this paper, we extended the GPR model
to the eTPR model. The latter inherits almost all the good features for the
GPR, and additionally it provides robust BLUP procedures in the presence of
outliers in both input and output spaces. Even with $m=1$, under some
kernels it gives robust prediction. Numerical studies show that the
eTPR is overall the best in prediction among the methods considered.

{  The GPR or eTPR model discussed in this paper is a concurrent functional regression model. The information consistency of the estimation of the unknown $f(\cdot)$ in equation (1) requires dense data, i.e. the sample size tends to infinity. We have shown that eTPR preforms robustly when there are outliers or the data are sparse in some areas. The idea has potential to be extended to a general functional data analysis framework, e.g. scalar-on-function or function-on-function regression model.  }

\appendix


\section{Properties of extended $t$-process}

Let $\mbox{\boldmath${\Sigma}$}$ be
an $n\times n$ symmetric and positive definite matrix, $\mbox{\boldmath${%
\mu}$}\in {R}{^{n}} $, $\nu >0$ and $\omega >0$. In this paper, $\mbox{\boldmath${Z}$}\sim
EMTD(\nu ,\omega ,\mbox{\boldmath${\mu}$},\mbox{\boldmath${\Sigma}$})$ means that a
random vector {$\mbox{\boldmath${Z}$}\in {R}{^{n}}$} has the
density function,
\begin{equation*}
p(z)=|2\pi \omega \mbox{\boldmath${\Sigma}$}|^{-1/2}\frac{\Gamma (n/2+\nu )}{%
\Gamma (\nu )}\left( 1+\frac{\mbox{{\boldmath${(z-\mu)}$}$^{T}$}%
\mbox{{\boldmath${\Sigma}$}$^{{-1}}$}\mbox{\boldmath${(z-\mu)}$}}{2\omega }%
\right) ^{-(n/2+\nu )},
\end{equation*}
where $\Gamma (\cdot )$ is the gamma function.


We may construct an EMTD via a double hierarchical generalized linear model
\citep{r9} as follows: \newline
\noindent \textbf{Lemma 1} \textit{If
\begin{equation*}
\mbox{\boldmath${Z}$}|r\sim N(\mbox{\boldmath${\mu}$},r%
\mbox{\boldmath
${\Sigma}$}),~~r\sim \mathrm{IG}(\nu ,\omega ),
\end{equation*}
where $\mathrm{IG}(\nu ,\omega )$ stands for an inverse gamma distribution
with the density function
\begin{equation*}
g(r)=\frac{1}{\Gamma (\nu )}(\frac{\omega }{r})^{\nu +1}\frac{1}{\omega }%
\exp {(-\frac{\omega }{r}),}
\end{equation*}
then, marginally $\mbox{\boldmath${Z }$}\sim EMTD(\nu ,\omega ,%
\mbox
{\boldmath${\mu}$},\mbox{\boldmath${\Sigma}$})$.}

\vskip 10pt \noindent \textbf{Proof}: From the construction
of $\mbox{\boldmath ${Z}$}$, we have
\begin{align}
p(\mbox{\boldmath ${z}$})&=\int_0^\infty p(\mbox{\boldmath ${z}$}|r)g(r)dr
\notag \\
&=\int_0^\infty |2\pi r\mbox{\boldmath ${\Sigma}$}|^{-1/2} \exp{(-\frac{(%
\mbox{\boldmath ${z}$}-\mbox{\boldmath ${\mu}$})^T%
\mbox{{\boldmath
${\Sigma}$}$^{-1}$}(\mbox{\boldmath ${z}$}-\mbox{\boldmath ${\mu}$})}{2r})}
\frac{1}{\Gamma(\nu)}(\frac{\omega}{r})^{\nu+1}\frac{1}{\omega}\exp{%
(-\omega/r)} dr  \notag \\
&=\int_0^\infty |2\pi \omega \mbox{\boldmath ${\Sigma}$}|^{-1/2} \frac{1}{%
\Gamma(\nu)} (\frac{\omega}{r})^{n/2+\nu-1} \exp{(-\frac{2\omega+(%
\mbox{\boldmath ${z}$}-\mbox{\boldmath ${\mu}$})^T%
\mbox{{\boldmath
${\Sigma}$}$^{-1}$}(\mbox{\boldmath ${z}$}-\mbox{\boldmath ${\mu}$})}{2r})} d%
\frac{\omega}{r}  \notag \\
&=|2\pi \omega \mbox{\boldmath ${\Sigma}$}|^{-1/2} \frac{\Gamma(n/2+\nu)}{%
\Gamma(\nu)}\left(1+\frac{(\mbox{\boldmath ${z}$}-\mbox{\boldmath ${\mu}$})^T%
\mbox{{\boldmath ${\Sigma}$}$^{-1}$}(\mbox{\boldmath ${z}$}-%
\mbox{\boldmath
${\mu}$})}{2\omega}\right)^{-(n/2+\nu)},  \notag
\end{align}
which is the density function of EMTD.$\sharp$

Properties of EMTD are as follows.

\noindent\textbf{Lemma 2} \textit{Let $\mbox{\boldmath${Z }$}\sim
EMTD(\nu,\omega,\mbox{\boldmath${\mu}$},\mbox{\boldmath${\Sigma}$}).$}

\begin{itemize}
\item[(i)]  \textit{If $\omega /\nu \rightarrow \lambda >0$ as $\nu
\rightarrow \infty $, then $\lim_{\nu \rightarrow \infty }EMTD(\nu ,\omega ,%
\mbox{\boldmath${\mu}$},\mbox{\boldmath${\Sigma}$})=N(\mu ,\lambda %
\mbox{\boldmath${\Sigma}$}).$ }

\item[(ii)]  \textit{{For any matrix $\mbox{\boldmath${A}$}\in R^{l\times n}$
with rank $l\leq n$, }} $\mathit{{\mbox{\boldmath${A }$}\mbox{\boldmath${Z}$}%
\sim EMTD(\nu ,\omega ,\mbox{\boldmath${A }$}\mu ,\mbox{\boldmath${A }$}%
\mbox{\boldmath${\Sigma}$}\mbox{{\boldmath${A}$}$^{T}$})}.\ }$

\item[(iii)]  \textit{Let $\mbox{\boldmath${Z}$}$ be partitioned as ${(%
\mbox{{\boldmath${Z}$}$_{1}^{T}$},\mbox{{\boldmath${Z}$}$_{2}^{T}$})}^{T}$
with lengths $n_{1}$ and $n_{2}=n-n_{1}$, and $\mbox{\boldmath${\mu}$}$ and $%
\mbox{\boldmath${\Sigma}$}$ have the corresponding partitions as $%
\mbox{\boldmath${\mu}$}=(\mbox{{\boldmath${\mu}$}$_{1}^{T}$},%
\mbox
{{\boldmath${\mu}$}$_{2}^{T}$})^{T}$ and $\mbox{\boldmath${\Sigma}$}=\left(
\begin{array}{cc}
\mbox{{\boldmath${\Sigma}$}$_{{11}}$} &
\mbox{{\boldmath
${\Sigma}$}$_{{12}}$} \\
\mbox{{\boldmath${\Sigma}$}$_{{12}}^{T}$} &
\mbox{{\boldmath
${\Sigma}$}$_{{22}}$}
\end{array}
\right) $. Then,}
\begin{align}
& \mathit{\mbox{{\boldmath${Z}$}$_{1}$}\sim EMTD(\nu ,\omega ,%
\mbox{{\boldmath${\mu}$}$_{1}$},\mbox{{\boldmath${\Sigma}$}$_{{11}}$}),}
\notag \\
& \mathit{\mbox{{\boldmath${Z}$}$_{2}$}|\mbox{{\boldmath${Z}$}$_{1}$}=%
\mbox{{\boldmath${z}$}$_{1}$}\sim EMTD(\nu ^{\ast },\omega ^{\ast },%
\mbox{\boldmath${\mu}$}^{\ast },\mbox{\boldmath${\Sigma}$}^{\ast })},  \notag
\end{align}
\textit{with $\nu ^{\ast }=n_{1}/2+\nu $, $\omega ^{\ast }=n_{1}/2+\omega $,
} $\mbox {{\boldmath${\mu}$}$^{*}$}=\mbox{{\boldmath${\Sigma}$}$_{{12}}^{T}$}%
\mbox{{\boldmath${\Sigma}$}$_{{11}}^{{-1}}$}(\mbox{{\boldmath${z}$}$_{1}$}-%
\mbox{{\boldmath${\mu}$}$_{1}$})+\mbox{{\boldmath${\mu}$}$_{2}$},$ $%
\mbox{{\boldmath${\Sigma}$}$^{*}$}=(2\omega +(\mbox{{\boldmath${z}$}$_{1}$}-%
\mbox{{\boldmath${\mu}$}$_{1}$})^{T}%
\mbox{{\boldmath
${\Sigma}$}$_{{11}}^{{-1}}$}(\mbox{{\boldmath${z}$}$_{1}$}-%
\mbox{{\boldmath
${\mu}$}$_{1}$}))\mbox{{\boldmath${\Sigma}$}$_{{22\cdot1}}$}/({2\omega +n_{1}%
}),$ \textit{and $\mbox{{\boldmath${\Sigma}$}$_{{22\cdot1}}$}=%
\mbox{{\boldmath
${\Sigma}$}$_{{22}}$}-\mbox{{\boldmath
${\Sigma}$}$_{{12}}^{T}$}\mbox{{\boldmath${\Sigma}$}$_{{11}}^{{-1}}$}%
\mbox{{\boldmath${\Sigma}$}$_{{12}}$}$. This gives } $E(\mbox{{%
\boldmath${Z}$}$_{2}$}|\mbox{{\boldmath${Z}$}$_{1}$})=%
\mbox
{{\boldmath${\mu}$}$^{*}$}\text{\textit{\ and }}Cov(\mbox{{%
\boldmath${Z}$}$_{2}$}|\mbox{{\boldmath${Z}$}$_{1}$})={\omega ^{\ast }}%
\mbox{{\boldmath${\Sigma}$}$^{*}$}/({\nu ^{\ast }-1}).$

\item[(iv)]  \textit{Let $r$ be a random effect in Lemma 1. Then, $r|%
\mbox{\boldmath${Z}$}\sim \mathrm{IG}(\tilde{\nu},\tilde{\omega})$ with $%
\tilde{\nu}=n/2+\nu ,$ $\tilde{\omega}=\omega +(\mbox{\boldmath${Z}$}-%
\mbox{\boldmath${\mu}$})^{T}\mbox{{\boldmath
${\Sigma}$}$^{{-1}}$}(\mbox{\boldmath${Z}$}-\mbox{\boldmath${\mu}$})/2$ and
\begin{align}
E(r|\mbox{\boldmath${Z}$})=\frac{2\omega +(\mbox{\boldmath${Z}$}-%
\mbox{\boldmath${\mu}$})^{T}\mbox{{\boldmath${\Sigma}$}$^{{-1}}$}(%
\mbox{\boldmath${Z}$}-\mbox{\boldmath${\mu}$})}{n+2\nu -2},  \notag \\
Var(r|\mbox{\boldmath${Z}$})=\frac{(2\omega +(\mbox{\boldmath${Z}$}-%
\mbox{\boldmath${\mu}$})^{T}\mbox{{\boldmath${\Sigma}$}$^{{-1}}$}(%
\mbox{\boldmath${Z}$}-\mbox{\boldmath${\mu}$}))^{2}}{(n+2\nu -2)^{2}(n/2+\nu
-2)}.  \notag
\end{align}
}
\end{itemize}

\vskip10pt \noindent \textbf{Proof}: The conclusions (i),
(ii) and $\mbox{{\boldmath ${Z}$}$_{1}$}\sim EMTD(\nu ,\omega ,%
\mbox{\boldmath ${\mu}$}_{1},\mbox{{\boldmath ${\Sigma}$}$_{11}$})$ in (iii)
are easily obtained by the definition of EMTD and Lemma 1. Now we only
prove that  $\mbox{{\boldmath ${Z}$}$_{2}$}|\mbox{{\boldmath ${Z}$}$_{1}$}\sim
EMTD(\nu ^{\ast },\omega ^{\ast },\mbox{{\boldmath ${\mu}$}$^{*}$},%
\mbox{{\boldmath ${\Sigma}$}$^{*}$})$. Let $a _{1}=(%
\mbox{{\boldmath
${z}$}$_{1}$}-\mbox{{\boldmath ${\mu}$}$_{1}$})^{T}%
\mbox{{\boldmath
${\Sigma}$}$_{11}^{-1}$}(\mbox{{\boldmath ${z}$}$_{1}$}-%
\mbox{{\boldmath
${\mu}$}$_{1}$})$ and $a _{2}=(\mbox{{\boldmath ${z}$}$_{2}$}-%
\mbox{{\boldmath ${\mu}$}$^{*}$})^{T}\mbox{{\boldmath ${\Sigma}$}$^{*-1}$}(%
\mbox{{\boldmath ${z}$}$_{2}$}-\mbox{{\boldmath ${\mu}$}$^{*}$})$, then $a
_{1}+a _{2}=(\mbox{\boldmath ${z}$}-\mbox{\boldmath ${\mu}$})^{T}%
\mbox{{\boldmath ${\Sigma}$}$^{-1}$}(\mbox{\boldmath ${z}$}-%
\mbox{\boldmath
${\mu}$})$. We have
\begin{align}
&p(\mbox{{\boldmath ${z}$}$_{2}$}|\mbox{{\boldmath ${z}$}$_{1}$}) =\frac{p(%
\mbox{\boldmath ${z}$})}{p(\mbox{{\boldmath ${z}$}$_{1}$})}  \notag \\
=&\frac{|2\pi \omega \mbox{\boldmath ${\Sigma}$}|^{-1/2}\frac{\Gamma
(n/2+\nu )}{\Gamma (\nu )}\left( 1+\frac{a_{1}+a_{2}}{2\omega }\right)
^{-(n/2+\nu )}}{|2\pi \omega \mbox{{\boldmath ${\Sigma}$}$_{11}$}|^{-1/2}%
\frac{\Gamma (n_{1}/2+\nu )}{\Gamma (\nu )}\left( 1+\frac{a_{1}}{2\omega }%
\right) ^{-(n_{1}/2+\nu )}}\propto \left( 1+\frac{a_{2}}{2\omega +a_{1}}%
\right) ^{-(n/2+\nu )},  \notag
\end{align}
which indicates $\mbox{{\boldmath ${Z}$}$_{2}$}|%
\mbox{{\boldmath
${Z}$}$_{1}$}\sim EMTD(\nu ^{\ast },\omega ^{\ast },%
\mbox{{\boldmath
${\mu}$}$^{*}$},\mbox{{\boldmath ${\Sigma}$}$^{*}$})$.

By combining definitions of IG and EMTD, we have
\begin{align}
& p(r|\mbox{\boldmath ${Z}$})=\frac{p(\mbox{\boldmath ${Z}$}|r)g(r)}{p(%
\mbox{\boldmath ${Z}$})}  \notag \\
& =\frac{1}{\Gamma (n/2+\nu )}\frac{1}{\omega +(\mbox{\boldmath ${Z}$}-%
\mbox{\boldmath ${\mu}$})^{T}\mbox{{\boldmath ${\Sigma}$}$^{-1}$}(%
\mbox{\boldmath ${Z}$}-\mbox{\boldmath ${\mu}$})/2}\left( \frac{\omega +(%
\mbox{\boldmath ${Z}$}-\mbox{\boldmath ${\mu}$})^{T}%
\mbox{{\boldmath
${\Sigma}$}$^{-1}$}(\mbox{\boldmath ${Z}$}-\mbox{\boldmath ${\mu}$})/2}{r}%
\right) ^{n/2+\nu +1}  \notag \\
& \exp {\left( -\frac{\omega +(\mbox{\boldmath ${Z}$}-%
\mbox{\boldmath
${\mu}$})^{T}\mbox{{\boldmath ${\Sigma}$}$^{-1}$}(\mbox{\boldmath ${Z}$}-%
\mbox{\boldmath ${\mu}$})/2}{r}\right) },  \notag
\end{align}
which indicates (iv) holds in this Lemma.$\sharp$

\vskip10pt \noindent \textbf{Proof of Proposition 1}: Proposition 1 can be
easily proved by using Lemma 2, so omitted here.$\sharp$

\section{Properties of eTPR model}

\subsection{Parameter estimation}

Let $\mbox{\boldmath ${\beta}$}=(\phi ,\mbox{\boldmath ${\theta}$}_1,...,\mbox{\boldmath ${\theta}$}_m)$, where $%
\phi $ is a parameter for $\epsilon (\mbox{\boldmath ${x}$})$ and $%
\mbox{\boldmath ${\theta}$}_i$ are those for $f_i(\mbox{{\boldmath ${x}$}})$ (parameter in the kernel $k_i$), $i=1,...,m$.
We know that $\mbox{{\boldmath ${y}$}$_{i}$}|\mbox{{\boldmath ${X}$}$_{i}$}\sim
EMTD(\nu ,\nu-1 ,0,\mbox{{\boldmath
${{\Sigma}}$}$_{in}$})$ with $\mbox{{\boldmath ${\Sigma}$}$_{in}$}=%
\mbox{{\boldmath
${K}$}$_{in}$}+\phi \mbox{{\boldmath ${I}$}$_{n}$}$, $\mbox{{\boldmath
${K}$}$_{in}$}=(k_{ijl})_{n\times n}$ and $k_{ijl}=k_i(%
\mbox{{\boldmath
${x}$}$_{ij}$},\mbox{{\boldmath${x}$}$_{il}$})$.
For given $\nu $, the marginal log-likelihood
of $\mbox{\boldmath ${\beta}$}$ is
\begin{align}
l(\mbox{\boldmath ${\beta}$};\nu)=& \sum_{i=1}^m\Big\{-\frac{n}{2}\log (2\pi (\nu-1) )-\frac{1%
}{2}\log |\mbox{{\boldmath ${{\Sigma}}$}$_{in}$}|{-(\frac{n}{2}+\nu )}%
\log \left( 1+\frac{S_i}{2(\nu-1) }\right)  \notag \\
& +\log (\Gamma (\frac{n}{2}+\nu ))-\log (\Gamma (\nu ))\Big\},  \notag
\end{align}
where $S_i=\mbox{{\boldmath ${y}$}$_{i}^{T}$}\mbox{{\boldmath
${{\Sigma}}$}$_{in}^{-1}$}\mbox{{\boldmath ${y}$}$_{i}$}$.
The score function of $\beta_k$, the $k$th element of $\mbox{\boldmath ${\beta}$}$, is
\begin{equation}
\frac{\partial l(\mbox{\boldmath ${\beta}$};\nu)}{\partial \beta_k}=\frac{1}{2}\sum_{i=1}^m Tr\left( \Big(s_{1i}\mbox{\boldmath
${\alpha}_i$}\mbox{{\boldmath
${\alpha}_i$}$^{T}$}-\mbox{{\boldmath ${{\Sigma}}$}$_{in}^{-1}$}\Big)%
\frac{\partial \mbox{{\boldmath ${{\Sigma}}$}$_{in}$}}{\partial \beta_k}\right) ,  \label{score-beta}
\end{equation}
where $\vesub{\alpha}{i}=%
\mbox{{\boldmath
${{\Sigma}}$}$_{in}^{-1}$}\mbox{{\boldmath ${y}$}$_{i}$}$, and $s_{1i}=({n+2\nu })/({2(\nu -1)+S_i})$.

A maximum likelihood (ML) estimate of $\mbox{\boldmath ${\beta}$}$ can be learned by using
gradient based methods, denoted by $\hve{\beta}$. The first derivative is given in (\ref{score-beta}) and the second derivative of $l(\mbox{\boldmath
${\beta}$};\nu)$ with respect to $\mbox{\boldmath ${\beta}$}$ is,
\begin{align}
&\frac{\partial^2 l(\mbox{\boldmath ${\beta}$};\nu)}{\partial \beta_k \partial \beta_k}=\frac{1}{2}\sum_{i=1}^m
Tr\left(\Big(s_{1i}\vesub{\alpha}{i}\vess{\alpha}{i}{T}-%
\mbox{{\boldmath
${{\Sigma}}$}$_{in}^{-1}$}\Big)\Big(\frac{\partial ^2
\mbox{{\boldmath
${{\Sigma}}$}$_{in}$}}{\partial\beta_k \partial \beta_k}-\frac{\partial
\mbox{{\boldmath
${{\Sigma}}$}$_{in}$}}{\partial\beta_k} %
\mbox{{\boldmath ${{\Sigma}}$}$_{in}^{-1}$} \frac{\partial %
\mbox{{\boldmath ${{\Sigma}}$}$_{in}$}}{\partial\beta_k}\Big)\right)  \notag \\
&-\frac{1}{2}Tr\left(s_{1i}\vesub{\alpha}{i}\vess{\alpha}{i}{T}\frac{\partial
\mbox{{\boldmath
${{\Sigma}}$}$_{in}$}}{\partial\beta_k} %
\mbox{{\boldmath ${{\Sigma}}$}$_{in}^{-1}$} \frac{\partial %
\mbox{{\boldmath ${{\Sigma}}$}$_{in}$}}{\partial\beta_k}\right)+\frac{1}{2}\frac{s_{1i}^2}{n+2\nu}\left\{Tr\left(%
\vesub{\alpha}{i}\vess{\alpha}{i}{T}\frac{\partial %
\mbox{{\boldmath ${{\Sigma}}$}$_{in}$}}{\partial\beta_k}\right)\right\}^2.  \label{sderivative}
\end{align}
Thus, variance of $\hat\beta_k$ can be estimated by using $\Big(-\frac{\partial^2 l(\mbox{\boldmath ${\beta}$};\nu)}{\partial \beta_k \partial \beta_k}\Big)^{-1}\Big|_{\ve\beta=\hve\beta}$, where $\hat\beta_k$  is the $k$th component of $\hve\beta$.

\subsection{Prediction}

The five covariance kernels are given as follows, \citep{r12,r16}, for $\ve{u},\ve{v}\in R^p$,
\begin{itemize}
\item Squared exponential kernel:
\begin{align}
k_{se}(\ve u,\ve v)=\eta_{0}\exp {\left( -\frac{1}{2}%
\sum_{l=1}^{p}\eta _{l}(u_{l}-v_{l})^{2}\right) },  \nonumber
\end{align}
where $\eta _{l}>0$, $l=0,1,...,p$.
\item Non-stationary linear kernel:
\begin{align}
k_{lin}(\ve u,\ve v)=\sum_{l=1}^{p}%
\eta_{l-1}u_{l}v_{l},  \nonumber
\end{align}
where $\eta_{l}>0$, $l=0,...,p-1$.
\item
von Mises-inspired kernel:
\begin{align}
k_{vm}(\ve u,\ve v)=\eta_{0}\exp {\left( \eta_1\Big(
\sum_{l=1}^{p}\cos(u_{l}-v_{l})-p\Big)\right) }, \nonumber
\end{align}
where $\eta_{0}>0$ and $\eta_1>0$.
\item Rational quadratic kernel:
\begin{align}
k_{rq}(\ve u,\ve v)=\left(1+(20^{1/\lambda}-1)\sum_{l=1}^{p}\eta _{l}(u_{l}-v_{l})^{2}\right)^{-\lambda},  \nonumber
\end{align}
where $\lambda>0$ and $\eta _{l}>0$, $l=1,...,p$.
\item Mat$\acute{e}$rn kernel: for a known $\alpha$,
\begin{align}
k_{m}(\ve u,\ve v)=\frac{1}{\Gamma(\alpha)2^{\alpha-1}}(\eta_1\|\ve u-\ve v\|)^\alpha
\mathcal{K}_{\alpha}(\eta_1\|\ve u-\ve v\|),  \nonumber
\end{align}
where $\eta_1>0$ and $\mathcal{K}_{\alpha}(\cdot)$ is a modified Bessel function of order $\alpha$.
When $\alpha=3/2$, $$k_{m}(\ve u,\ve v)=(1+\eta_1(\sum_{l=1}^p(u_l-v_l)^2)^{1/2})\exp{(-\eta_1(\sum_{l=1}^p(u_l-v_l)^2)^{1/2})}.$$
\end{itemize}

\vskip10pt \noindent \textbf{Proof of Proposition 2}:
We show that under the kernel functions $k_{se}$, $k_{lin}$ and $k_{vm}$, when $\eta_0$ is unknown and estimated, the predictions of $f_1(\mbox{{\boldmath ${X}$}$_{1}$})$ and $f_1(\mbox{\boldmath${u}$})$ from eTPR models have the same values as those from GPR models.
When $\eta_0$ is a constant such as $\eta_0=1$, eTPR models under each  of above kernels have different predictions to those
from GPR models. We take the squared exponential kernel and the rational quadratic kernel as examples to prove the properties.


Under the squared exponential kernel $k_1=k_{se}$, we can reparametrize the kernel function $k_1$ as ${k}_1^*=k_1/\phi$, that is
\begin{align}
{k}_1^*(\ve u,\ve v)
={\eta}^*_{0}\exp {\left( -\frac{1}{2}%
\sum_{l=1}^{p}\eta _{l}(u_{l}-v_{l})^{2}\right) },  \nonumber
\end{align}
where  ${\eta}^*_{0}=\eta _{0}/\phi$. Thus, we can rewrite
\begin{equation}
\left(
\begin{array}{c}
f_1 \\
\epsilon_1
\end{array}
\right) \sim ETP\left( \nu ,\nu-1 ,\left(
\begin{array}{c}
0 \\
0
\end{array}
\right) ,\phi\left(
\begin{array}{cc}
{k}_1^* & 0 \\
0 & I_1
\end{array}
\right) \right) ,  \nonumber
\end{equation}
where $I_1(\mbox{\boldmath ${u}$},\mbox{\boldmath ${v}$}%
)=I(\mbox{\boldmath ${u}$}=\mbox{\boldmath ${v}$})$.

Let $\mbox{\boldmath ${\theta}$}^*_{1}=\{\eta^*_{0},\eta_{l},l=1,...,p\}$, and $\mbox{\boldmath ${\beta}$}^*=(\phi ,\vess{\theta}{1}{*T})^T$.
Under this reparametrization, for a new data point $\ve x=\ve u$,
the prediction of $f_1(\mbox{{\boldmath ${u}$}})$ becomes
\begin{align}
E(f_1(\mbox{{\boldmath ${u}$}})|%
\mbox{{\boldmath ${\cal D}$}$_{n}$})={\ve k}_{1u}
\mbox{{\boldmath ${\Sigma}$}$_{1n}^{-1}$}\mbox{{\boldmath${y}$}}_1={\ve k}^*_{1u}\vess{H}{1n}{-1}\mbox{{\boldmath${y}$}}_1,
\nonumber
\end{align}
where ${\ve k}_{1u}=(k_1(\vesub{x}{11},\ve u),...,k_1(\vesub{x}{1n},\ve u))^T$, ${\ve k}^*_{1u}=({k}_1^*(\vesub{x}{11},\ve u),...,{k}_1^*(\vesub{x}{1n},\ve u))^T$,
${\ve K}^*_{1n}=({k}_1^*(\vesub{x}{1i},\vesub{x}{1j}))_{n\times n}$ and $\vesub{H}{1n}={\ve K}^*_{1n}+\vesub{I}{n}$.
We can see this prediction only depends on ${\eta}^*_{0}$ and ${\eta}_{l}$. The predictive covariance under the eTPR model is
\begin{align}
& Var(f_1(\mbox{{\boldmath ${u}$}})|\mbox{{\boldmath ${\cal D}$}$_{n}$})=s_{01}(k_1(\ve u,\ve u)-{\ve k}_{1u}^T\vess{\Sigma}{1n}{-1}{\ve k}_{1u})
=s_{01}\phi({k}_1^*(\ve u,\ve u)-{{\ve k}}_{1u}^{*T}\vess{H}{1n}{-1}{{\ve k}}^*_{1u}),  \notag \\
& s_{01}=\frac{\vess{y}{1}{T}\mbox{{\boldmath
${\Sigma}$}$_{1n}^{-1}$}\mbox{{\boldmath
${y}$}}_1+2(\nu -1)}{n+2(\nu -1)}=\frac{\phi^{-1}\vess{y}{1}{T}\mbox{{\boldmath
${H}$}$_{1n}^{-1}$}\mbox{{\boldmath
${y}$}}_1+2(\nu -1)}{n+2(\nu -1)},  \notag
\end{align}
which indicates that it depends on ${\eta}^*_{0}$, ${\eta}_{l}$, $\phi$ and $\nu$.

From (\ref{score-beta}), letting the score function of $\phi$ be 0, we obtain
\begin{align}
s_{11} \phi^{-1}=\frac{n}{\vess{y}{1}{T}\vess{H}{1n}{-1}\mbox{{\boldmath ${y}$}}_1}.\nonumber
\end{align}
It gives a ML estimator of $\phi$ as
\begin{align}
\hat\phi=\frac{\nu}{\nu-1}\frac{\mbox{{\boldmath ${y}$}$_{1}^{T}$}\vess{H}{1n}{-1}\mbox{{\boldmath ${y}$}$_{1}$}}{n}=\frac{\nu}{\nu-1}\tilde{\phi},\nonumber
\end{align}
where $\tilde{\phi}$ is ML estimator of $\phi$ under the GPR model ($s_{11}=1$).

For $\eta_{l}$ and ${\eta}^*_{0}$, we have their score equations,
\begin{align}
\frac{\partial l(\mbox{\boldmath ${\beta}$}^*;\nu)}{\partial \eta_{l}}=\frac{\partial l(\mbox{\boldmath ${\beta}$};\nu)}{\partial \eta_{l}}
=&\frac{1}{2} Tr\left( \Big(s_{11}\phi^{-1}\vess{H}{1n}{-1}\vesub{y}{1}\vess{y}{1}{T}\vess{H}{1n}{-1}-\mbox{{\boldmath ${H}$}$_{1n}^{-1}$}\Big)%
\frac{\partial \mbox{{\boldmath ${H}$}$_{1n}$}}{\partial \eta_{l}}\right)=0,\nonumber\\ 
\frac{\partial l(\mbox{\boldmath ${\beta}$}^*;\nu)}{\partial {\eta}^*_{0}}=\frac{\partial l(\mbox{\boldmath ${\beta}$};\nu)}{\partial {\eta}^*_{0}}
=&\frac{1}{2} Tr\left( \Big(s_{11}\phi^{-1}\vess{H}{1n}{-1}\vesub{y}{1}\vess{y}{1}{T}\vess{H}{1n}{-1}-\mbox{{\boldmath ${H}$}$_{1n}^{-1}$}\Big)%
\frac{\partial \mbox{{\boldmath ${H}$}$_{1n}$}}{\partial {\eta}^*_{0}}\right)=0.\nonumber 
\end{align}
Thus, we have
\begin{align}
\frac{\partial l(\mbox{\boldmath ${\beta}$}^*;\nu)}{\partial \eta_{l}}=\frac{1}{2} Tr\left( \Big(\frac{n}{\vess{y}{1}{T}\vess{H}{1n}{-1}\mbox{{\boldmath ${y}$}}_1}\vess{H}{1n}{-1}\vesub{y}{1}\vess{y}{1}{T}\vess{H}{1n}{-1}-\mbox{{\boldmath ${H}$}$_{1n}^{-1}$}\Big)%
\frac{\partial \mbox{{\boldmath ${H}$}$_{1n}$}}{\partial \eta_{l}}\right)=0,\label{rescore-eta-m1}\\
\frac{\partial l(\mbox{\boldmath ${\beta}$}^*;\nu)}{\partial {\eta}^*_{0}}=\frac{1}{2} Tr\left( \Big(\frac{n}{\vess{y}{1}{T}\vess{H}{1n}{-1}\mbox{{\boldmath ${y}$}}_1}\vess{H}{1n}{-1}\vesub{y}{1}\vess{y}{1}{T}\vess{H}{1n}{-1}-\mbox{{\boldmath ${H}$}$_{1n}^{-1}$}\Big)%
\frac{\partial \mbox{{\boldmath ${H}$}$_{1n}$}}{\partial {\eta}^*_{0}}\right)=0.\label{rescore-ratio-m1}
\end{align}
We can see that the score equations (\ref{rescore-eta-m1}) and (\ref{rescore-ratio-m1}) for $\eta_{l}$ and ${\eta}^*_{0}$ do not depend on $\nu$ and $s_{11}$, and they are the same as those under the GPR model. Therefore, the parameters ${\eta}^*_{0}$, $\eta_{l}$ under the eTPR model are estimated with the same values as those under the GPR model, which leads to the same prediction of $f_1(\ve u)$.

Plugging $\hat\phi$ in $Var(f_1(\mbox{{\boldmath ${u}$}})|\mbox{{\boldmath ${\cal D}$}$_{n}$})$, we have
\begin{align}
Var(f_1(\mbox{{\boldmath ${u}$}})|\mbox{{\boldmath ${\cal D}$}$_{n}$})=&\frac{n+2\nu}{n+2(\nu-1)}\tilde{\phi}({k}_1^*(\ve u,\ve u)-{{\ve k}}_{1u}^{*T}\vess{H}{1n}{-1}{{\ve k}}_{1u}^*)\nonumber\\
=&\frac{n+2\nu}{n+2(\nu-1)}\widetilde{Var}(f_1(\mbox{{\boldmath ${u}$}})|\mbox{{\boldmath ${\cal D}$}$_{n}$}),\nonumber
\end{align}
where $\widetilde{Var}(f_1(\mbox{{\boldmath ${u}$}})|\mbox{{\boldmath ${\cal D}$}$_{n}$})$ is conditional variance of $f_1(\mbox{{\boldmath ${u}$}})$ under GPR model.
It follows that the eTPR has slightly bigger variance estimate of the predictor than the GPR.


Under the rational quadratic kernel $k_1=k_{rq}$, the score equations for $\phi$, $\lambda$ and $\eta_{l}$ are

\begin{align}
&\frac{\partial l(\mbox{\boldmath ${\beta}$};\nu)}{\partial \phi}=\frac{1}{2}Tr\left( s_{11}\vesub{\alpha}{1}\vess{\alpha}{1}{T}-\mbox{{\boldmath ${{\Sigma}}$}$_{1n}^{-1}$}\right)=0,\nonumber\\
&\frac{\partial l(\mbox{\boldmath ${\beta}$};\nu)}{\partial \lambda}=\frac{1}{2}Tr\left( \Big(s_{11}\vesub{\alpha}{1}\vess{\alpha}{1}{T}-\mbox{{\boldmath ${{\Sigma}}$}$_{1n}^{-1}$}\Big)\frac{\partial \mbox{{\boldmath ${{\Sigma}}$}$_{1n}$}}{\partial \lambda}\right)=0,\nonumber\\
&\frac{\partial l(\mbox{\boldmath ${\beta}$};\nu)}{\partial \eta_{l}}=\frac{1}{2}Tr\left( \Big(s_{11}\vesub{\alpha}{1}\vess{\alpha}{1}{T}-\mbox{{\boldmath ${{\Sigma}}$}$_{1n}^{-1}$}\Big)\frac{\partial \mbox{{\boldmath ${{\Sigma}}$}$_{1n}$}}{\partial \eta_{l}}\right)=0,\nonumber
\end{align}
which are different from those under the GPR model. Hence, the eTPR model has the different estimate value of $\ve\beta$ from the GPR model.

Note that
\begin{equation*}
s_{01}=\frac{\mbox{{\boldmath
${y}$}$_{1}^{T}$}\mbox{{\boldmath
${\Sigma}$}$_{1n}^{-1}$}\mbox{{\boldmath
${y}$}$_{1}$}+2(\nu -1)}{n+2(\nu -1)}=\frac{(\mbox{{\boldmath ${y}$}$_{1}$}-%
\hat{\mbox{\boldmath ${f}$}}_{1n})^{T}\mbox{{\boldmath${
\Sigma}$}$_{1n}$}(\mbox{{\boldmath ${y}$}$_{1}$}-\hat{\mbox{\boldmath ${f}$}}%
_{1n})/\phi ^{2}+2(\nu -1)}{n+2(\nu -1)},
\end{equation*}
where $\hat{\mbox{\boldmath ${f}$}}_{1n}=\mbox{{\boldmath ${\mu}$}$_{1n}$}$
is the prediction for $f_1(\mbox{{\boldmath ${X}$}$_{1}$})$. Thus, the predictive variance under the eTPR model decreases if the model
fits the responses $\mbox{{\boldmath ${y}$}$_{i}$}$ better while that under
the GPR model is still independent of the model fit.$\sharp$

\vskip10pt \noindent \textbf{Proof of Proposition 3}:

When $m>1$ such as $m=2$,  we use combination of $k_i=k_{se}+k_{lin}$ as an example to illustrate our methods, that is
\begin{align}
k_i(\ve u,\ve v)=k(\ve u,\ve v;%
\mbox{\boldmath ${\theta}_i$})
=\eta _{i0}\exp {\left( -\frac{1}{2}%
\sum_{l=1}^{p}\eta _{il}(u_{l}-v_{l})^{2}\right) }+\sum_{l=1}^{p}%
\xi_{il}u_{l}v_{l},  \label{kerfun}
\end{align}
where $\mbox{\boldmath ${\theta}$}_i=\{\eta _{i0},\eta_{il},\xi
_{il},l=1,...,p\}$ are a set of parameters.
Let parameter $\ve\beta=(\phi,\vess{\theta}{1}{T},\vess{\theta}{2}{T})^T$.
The predictions from both GPR and eTPR are exactly the same under this kernel when $m=1$.

From (\ref{score-beta}), similar to the case with $m=1$, the score  equation of $\phi$ becomes
\begin{align}
\phi=\frac{1}{2n}\sum_{i=1}^2 s_{1i}\mbox{{\boldmath ${y}$}$_{i}^{T}$}\vess{H}{in}{-1}\mbox{{\boldmath ${y}$}$_{i}$},\label{rescore-phi-1}
\end{align}
where $\vesub{H}{in}=\vesub{K}{in}/\phi+I_n$.
For other parameters, such as $\beta_3=\eta_{11}$, we have its score equation,
\begin{align}
\frac{\partial l(\mbox{\boldmath ${\beta}$};\nu)}{\partial \eta_{11}}
=&\frac{1}{2} Tr\left( \Big(s_{11}\phi^{-1}\vess{H}{1n}{-1}\vesub{y}{1}\vess{y}{1}{T}\vess{H}{1n}{-1}-\mbox{{\boldmath ${H}$}$_{1n}^{-1}$}\Big)%
\frac{\partial \mbox{{\boldmath ${H}$}$_{1n}$}}{\partial \eta_{11}}\right)=0,\label{rescore-eta}
\end{align}

From (\ref{rescore-phi-1}) and (\ref{rescore-eta}), we can see that the score equation for $\eta_{11}$ depend on $\nu$, $s_{11}$ and $s_{12}$. So they are different under the eTPR model from those under the GPR model.
Thus, the predictions of $f_1(\mbox{{\boldmath ${X}$}$_{1}$})$ and $f_1(\ve u)$ have different values for the eTPR and GPR models.

When $m>1$, we may estimate $\nu$ by using the following derivative,
\begin{align}  \label{score-df}
\frac{\partial l(\ve\beta;\nu)}{\partial \nu}=&-\frac{1}{2}\sum_{i=1}^m\Big\{%
\frac{n}{\nu-1}+2\log(1+\frac{S_i}{2(\nu-1)})- \frac{(n+2\nu)S_i}{%
2(\nu-1)^2+(\nu-1) S_i}\nonumber\\
&-2\psi(\frac{n}{2}+\nu)+2\psi(\nu)\Big\},
\end{align}
where $\psi(\cdot)$ is digamma function satisfying $\psi(x+1)=\psi(x)+1/x$. $\sharp$

\vskip10pt \noindent \textbf{Variance of prediction $\hat y_i(%
\mbox{\boldmath ${u}$})$}:

From the hierarchical sampling method in Lemma 1, we have
\begin{equation*}
\left(
\begin{array}{c}
f_i \\
\epsilon_i
\end{array}
\right)\Bigg|r_i \sim GP\left( \nu,(\nu-1),0 ,\left(
\begin{array}{cc}
r_ik_i & 0 \\
0 & r_ik_{\epsilon}
\end{array}
\right) \right),~~r_i\sim \mathrm{IG}(\nu ,(\nu-1) ),
\end{equation*}
which suggests that conditional distribution of $\mbox{{\boldmath ${y}$}$_{i}$}|f_i,r_i,%
\mbox{{\boldmath ${X}$}$_{i}$}\sim N(f_i(\mbox{{\boldmath ${X}$}$_{i}$}),r_i\phi %
\mbox{{\boldmath ${I}$}$_{n}$})$ and conditional distribution of $%
\mbox{{\boldmath ${y}$}$_{i}$}|r_i,\mbox{{\boldmath ${X}$}$_{i}$}\sim N(0,r_i\mbox{{\boldmath ${\Sigma}$}$_{in}$})$.
For given $r_i$, it follows that $E(\hat y_i(%
\mbox{\boldmath
${u}$})|r_i,\mbox{{\boldmath ${\mathcal{D}}$}$_{n}$})=%
\mbox{{\boldmath
${k}$}$_{{iu}}^{T}$}\mbox{{\boldmath
${{\Sigma}}$}$_{in}^{{-1}}$}\mbox{{\boldmath${ y}$}$_{i}$}$ and $Var(\hat y_i(%
\mbox{\boldmath ${u}$})|r_i,\mbox{{\boldmath
${\mathcal{D}}$}$_{n}$})=r_i(k_i(\mbox{\boldmath ${u}$},\mbox{\boldmath ${u}$})-%
\mbox{{\boldmath ${k}$}$_{iu}^{T}$}\mbox{{\boldmath${{%
\Sigma}}$}$_{in}^{-1}$}\mbox{{\boldmath ${k}$}$_{i u}$}+\phi)$.
Consequently, we have
\begin{align}
&Var(\hat y_i(\mbox{\boldmath ${u}$})|\mbox{{\boldmath
${\mathcal{D}}$}$_{n}$})=E((\hat y_i(\mbox{\boldmath ${u}$}))^2|%
\mbox{{\boldmath ${\mathcal{D}}$}$_{n}$})- (E(\hat y_i(\ve u)|\mbox{{\boldmath ${\mathcal{D}}$}$_{n}$}))^2  \notag \\
=& E_{r_i}[\{Var(\hat y_i(\mbox{\boldmath ${u}$})|r_i,%
\mbox{{\boldmath
${\mathcal{D}}$}$_{n}$})+(E(\hat y_i(\mbox{\boldmath ${u}$})|r_i,%
\mbox{{\boldmath ${\mathcal{D}}$}$_{n}$}))^2\}|%
\mbox{{\boldmath
${\mathcal{D}}$}$_{n}$}]- (E(\hat y_i(\mbox{\boldmath ${u}$})|%
\mbox{{\boldmath
${\mathcal{D}}$}$_{n}$}))^2  \notag \\
=&E_{r_i}(Var(\hat y_i(\mbox{\boldmath ${u}$})|r_i,%
\mbox{{\boldmath
${\mathcal{D}}$}$_{n}$})|\mbox{{\boldmath ${\mathcal{D}}$}$_{n}$})+(E(\hat y_i(%
\mbox{\boldmath ${u}$})|\mbox{{\boldmath ${\mathcal{D}}$}$_{n}$}))^2-
(E(\hat y_i(\mbox{\boldmath ${u}$})|\mbox{{\boldmath
${\mathcal{D}}$}$_{n}$}))^2  \notag \\
=&s_{0i}\Big(k_i(\mbox{\boldmath ${u}$},\mbox{\boldmath ${u}$})-%
\mbox{{\boldmath
${k}$}$_{i u}^{T}$}\mbox{{\boldmath${{\Sigma}}$}$_{in}^{-1}$}%
\mbox{{\boldmath ${k}$}$_{i
u}$}+\phi\Big),  \notag
\end{align}
where $s_{0i}=E(r_i|\vesub{\mathcal D}{n})={(2\nu-2+\mbox{{\boldmath
${ y}$}$_{i}$}^{T}%
\mbox{{\boldmath
${{\Sigma}}$}$_{in}^{-1}$}\mbox{{\boldmath${ y}$}$_{i}$}}/{(n+2\nu-2)}$.

\vskip 10pt
\section{Robustness and consistency}
Since kernel functions $k_i$ depend on parameters $\vesub{\theta}{i}$, from now on let $k_i(\ve u,\ve v)=k_i(\ve u,\ve v; \vesub{\theta}{i})$ for
convenient description.

\noindent \textbf{Proof of Proposition 4}:
From (\ref{score-beta}), the score function of $\beta_k$ from the eTPR model is 
\begin{align}
&s_{k}(\mbox{\boldmath ${\beta}$};\mbox{{\boldmath ${y}$}$_{1}$},...,\vesub{y}{m})=\frac{1}{2}\sum_{i=1}^m
Tr\left(\Big(s_{1i}\mbox{{\boldmath ${{\Sigma}}$}$_{in}^{-1}$}%
\mbox{{\boldmath ${y}$}$_{i}$}\mbox{{\boldmath ${ y}$}$_{i}^{T}$}%
\mbox{{\boldmath ${{\Sigma}}$}$_{in}^{-1}$}-%
\mbox{{\boldmath
${{\Sigma}}$}$_{in}^{-1}$}\Big)\frac{\partial
\mbox{{\boldmath
${{\Sigma}}$}$_{in}$}}{\partial\beta_k}\right).
\notag
\end{align}
Let $s_{T}(\mbox{\boldmath ${\beta}$};\mbox{{\boldmath
${y}$}$_{1}$},...,\vesub{y}{m})=(s_{1}(\mbox{\boldmath ${\beta}$};\mbox{{\boldmath ${y}$}$_{1}$},...,\vesub{y}{m}),\cdots,s_{L}(\mbox{\boldmath ${\beta}$};\mbox{{\boldmath ${y}$}$_{1}$},...,\vesub{y}{m}))^T$, where $L$ is length of $\ve\beta$.
When $s_{1i}=1$, the score function becomes that under the GPR model.
The term $s_{1i}=({n+2\nu})/({2(\nu-1)+\mbox{{\boldmath ${y}$}$_{i}^{T}$}%
\mbox{{\boldmath ${{\Sigma}}$}$_{in}^{-1}$}%
\mbox{{\boldmath
${y}$}$_{i}$}})$ in $s_{T}(\mbox{\boldmath ${\beta}$};\mbox{{\boldmath
${y}$}$_{1}$},...,\vesub{y}{m})$ plays an important role in estimating $%
\mbox{\boldmath ${\beta}$}$. For example, when $y_{ij}\rightarrow \infty$ for some $j$, the score $%
s_{T}(\mbox{\boldmath ${\beta}$};\vesub{y}{1},...,\vesub{y}{m})$ is bounded, while that from the GPR model
tends to $\infty$.

Let $T(F_n)=T_n(y_{11},...,y_{mn})$
be an estimate of $\mbox{\boldmath ${\beta}$}$, where $F_n$ is the empirical
distribution of $\{y_{11},...,y_{mn}\}$ and $T$ is a functional on some subset of
all distributions. Influence function of $T$ at $F$ (Hampel \textit{et al}., 1986) is
defined as
\begin{align}
IF(\ve y; T, F)=\lim_{t\rightarrow 0}\frac{T((1-t)F+t \delta_{\ve y})-T(F)}{t},  \notag
\end{align}
where $\delta_{\ve y}$ put mass 1 on point $\ve y$ and 0 on others.

For given parameter $\nu$, following \cite{r25} estimator $\hat{%
\mbox{\boldmath ${\beta}$}}$ of $\mbox{\boldmath ${\beta}$}$ has the
influence function
\begin{align}
IF(\ve y; \hat{\mbox{\boldmath ${\beta}$}}, F)=-\left(E\left(\frac{\partial^2 l(%
\mbox{\boldmath ${\beta}$};\nu)}{\partial
\mbox{\boldmath ${\beta
}$} \partial \mbox{{\boldmath ${\beta}$}$^{T}$}}\right)\right)^{-1}s_{T}(%
\mbox{\boldmath ${\beta}$};\ve y).  \notag
\end{align}
Note that the matrix ${\partial^2 l(\mbox{\boldmath
${\beta}$};\nu)}/{\partial \mbox{\boldmath ${\beta }$} \partial %
\mbox{{\boldmath ${\beta}$}$^{T}$}}$ is bounded according to $%
\mbox{{\boldmath ${y}$}$_{i}$}$, $i=1,...,m$, which indicates that the influence function
of $\hat{\mbox{\boldmath ${\beta}$}}$ is bounded under the eTPR model.
Similarly, we can obtain that the score function ($s_{1i}=1$) under the GPR model is unbound,
which leads to unbound influence function of parameter estimate.$\sharp$

\vskip 10pt \noindent \textbf{Proof of the equation (\ref{bspred})}: From  Bayes' Theorem, we have
\begin{align}
&\prod_{l=1}^np_{\phi_0,\theta_i}(y_{il} |\mbox{{\boldmath ${X}$}$_{il}$}, %
\mbox{{\boldmath ${y}$}$_{i(l-1)}$}) =p_{\phi_0,\theta_i}(y_{i1} |\mbox{{\boldmath
${X}$}$_{i1}$}) \prod_{l=2}^n\int_{{\mathcal{F}}} p_{\phi_0}(y_{il} |f,%
\mbox{{\boldmath
${X}$}$_{il}$}, \mbox{{\boldmath ${y}$}$_{i(l-1)}$})dp_{\theta_i}(f|%
\mbox{{\boldmath
${X}$}$_{il}$}, \mbox{{\boldmath
${y}$}$_{i(l-1)}$})  \notag \\
=&p_{\phi_0,\theta_i}(y_{i1} |\mbox{{\boldmath ${X}$}$_{i1}$})
\prod_{l=2}^n\int_{{\mathcal{F}}} \frac{p_{\phi_0}(%
\mbox{{\boldmath
${y}$}$_{il}$}|f,\mbox{{\boldmath ${X}$}$_{il}$})dp_{\theta_i}(f)}{\int_{{%
\mathcal{F}}} p_{\phi_0}(\mbox{{\boldmath ${y}$}$_{i(l-1)}$}|f{^{\prime}},%
\mbox{{\boldmath
${X}$}$_{i(l-1)}$})dp_{\theta_i}(f{^{\prime}})}  \notag \\
=&{\int_{{\mathcal{F}}} p_{\phi_0}(\mbox{{\boldmath ${y}$}$_{i}$}|f, %
\mbox{{\boldmath ${X}$}$_{i}$})dp_{\theta_i}(f)}=p_{\phi_0,\theta_i}(%
\mbox{{\boldmath ${y}$}$_{i}$} |\mbox{{\boldmath ${X}$}$_{i}$}),  \notag
\end{align}
which shows that the equation (\ref{bspred}) holds.$\sharp$

\vskip 10pt \noindent \textbf{Lemma 3} \textit{Suppose $%
\mbox{{\boldmath
${y}$}$_{i}$}=\{y_{i1},...,y_{in}\}$ are generated from the eTPR model (\ref{assum1}%
) with the mean function $h(\mbox{\boldmath ${x}$})=0$, and covariance
kernel function $k_i$ is bounded and continuous in parameter $%
\mbox{\boldmath
${\theta}$}_i $. It also assumes that the estimate $\hat{%
\mbox{\boldmath
${\beta}$}}$ almost surely converges to $\mbox{\boldmath ${\beta}$}$ as $%
n\rightarrow \infty$. Then for a positive constant $c$, and any $\varepsilon>0$%
, when $n$ is large enough, we have
\begin{align}
&\frac{1}{n}(-\log p_{\phi_0,\hat\theta_i}(\mbox{{\boldmath ${y}$}$_{i}$}|%
\mbox{{\boldmath ${X}$}$_{i}$})+\log p_{\phi_0}( \mbox{{\boldmath
${y}$}$_{i}$}|f_{0i},\mbox{{\boldmath ${X}$}$_{i}$}))  \notag \\
\leq& \frac{1}{n}\left\{\frac{1}{2}\log|\mbox{{\boldmath ${I}$}$_{n}$}%
+\phi_0^{-1} \mbox{{\boldmath ${K}$}$_{in}$}|+ \frac{q_i^2+2(\nu-1)}{2(n+2\nu-2)%
}(||f_{0i}||^2_k+c)+c\right\}+ \varepsilon,  \notag
\end{align}
where $\mbox{{\boldmath ${K}$}$_{in}$}=(k_i(\vesub{x}{ij},\vesub{x}{il}))_{n\times n}$, $q_i^2=(%
\mbox{{\boldmath ${y}$}$_{i}$}-f_{0i}(\mbox{{\boldmath ${X}$}$_{i}$}))^T(%
\mbox{{\boldmath ${y}$}$_{i}$}-f_{0i}(\mbox{{\boldmath ${X}$}$_{i}$}))/\phi_0$,
$\mbox{{\boldmath ${I}$}$_{n}$}$ is the $n\times n$ identity matrix, and $%
||f_{0i}||_k$ is the reproducing kernel Hilbert space norm of $f_{0i}$ associated
with kernel function $k_i(\cdot,\cdot;\vesub{\theta}{i})$. }

\vskip 10pt \textbf{Proof}: From Proposition 1, it follows that there exists a
variable $r_i\sim IG(\nu,(\nu-1))$, conditional
on $r_i$ we have
\begin{align}
\left(
\begin{array}{c}
f_i \\
\epsilon_i
\end{array}
\right)\Big|r_i\sim GP\left(\left(
\begin{array}{c}
0 \\
0
\end{array}
\right),\left(
\begin{array}{cc}
r_i k_i & 0 \\
0 & r_i k_{\epsilon}
\end{array}
\right)\right),  \notag
\end{align}
where $GP(h,k)$ stands for Gaussian process with mean function $h$ and
covariance function $k$. Then conditional on $r_i$, the extended t-process
regression model (\ref{assumed}) becomes Gaussian process regression model
\begin{eqnarray}  \label{rGPR}
y_i(\mbox{\boldmath ${x}$})=\tilde{f}_i(\mbox{\boldmath ${x}$})+\tilde{\epsilon}_i(%
\mbox{\boldmath ${x}$}),
\end{eqnarray}
where $\tilde{f}_i=f_i|r_i\sim GP(0, r_ik_i(\cdot,\cdot;\theta_i))$, $\tilde{\epsilon}_i%
|r_i\sim GP(0,r_ik_{\epsilon}(\cdot,\cdot;\phi_0))$ , and $\tilde{f}_i $ and error
term $\tilde{\epsilon}_i$ are independent. Denoted $\tilde{p}$ by computation of conditional probability density
for given $r_i$. Based on the model (\ref{rGPR}), let
\begin{align}
&p_{G}(\mbox{{\boldmath ${y}$}$_{i}$}|r_i,\mbox{{\boldmath ${X}$}$_{i}$}%
)=\int_{\mathcal{F}}{p}_{\phi_0}(\mbox{{\boldmath ${y}$}$_{i}$}|\tilde{%
f},r_i,\mbox{{\boldmath ${X}$}$_{i}$}) d\tilde{p}_{\theta_i}(\tilde{f}),  \notag \\
&p_{0}(\mbox{{\boldmath ${y}$}$_{i}$}|r_i,\mbox{{\boldmath ${X}$}$_{i}$})=%
{p}_{\phi_0}(\mbox{{\boldmath
${y}$}$_{i}$}|f_{0i},r_i,\mbox{{\boldmath ${X}$}$_{i}$}),  \notag
\end{align}
where $\tilde{p}_{\theta_i}$ is the induced measure from Gaussian process $%
GP(0,r_ik_i(\cdot,\cdot;\hat{\mbox{\boldmath ${\theta}$}_i}))$.

We know that variable $r_i$ is independent of covariates $%
\mbox{{\boldmath ${X}$}$_{i}$}$. Then it easily shows that
\begin{align}
&p_{\phi_0,\hat{\theta}_i}(\mbox{{\boldmath ${y}$}$_{i}$}|%
\mbox{{\boldmath
${X}$}$_{i}$})=\int p_{G}(\mbox{{\boldmath
${y}$}$_{i}$}|r,\mbox{{\boldmath ${X}$}$_{i}$}) g(r) dr,  \label{tpr1} \\
&p_{\phi_0}(\mbox{{\boldmath
${y}$}$_{i}$}|f_{0i},\mbox{{\boldmath ${X}$}$_{i}$})=\int p_{0}(%
\mbox{{\boldmath
${y}$}$_{i}$}|r,\mbox{{\boldmath ${X}$}$_{i}$}) g(r) dr.  \label{tpr0}
\end{align}

Suppose that for any given $r_i$, we have
\begin{align}  \label{GPcons}
&-\log p_{G}(\mbox{{\boldmath ${y}$}$_{i}$}|r_i,\mbox{{\boldmath ${X}$}$_{i}$}%
)+\log p_{0}(\mbox{{\boldmath ${y}$}$_{i}$}|r_i,\mbox{{\boldmath ${X}$}$_{i}$})
\notag \\
\leq& \frac{1}{2}\log|\mbox{{\boldmath ${I}$}$_{n}$}+\phi_0^{-1} %
\mbox{{\boldmath ${K}$}$_{in}$}|+ \frac{r_i}{2}(||f_{0i}||^2_k+c)+c+n\varepsilon.
\end{align}
Then we have
\begin{align}  \label{loggp}
&-\log\int p_{G}(\mbox{{\boldmath ${y}$}$_{i}$}|r,%
\mbox{{\boldmath
${X}$}$_{i}$}) g(r) dr \leq \frac{1}{2}\log|\mbox{{\boldmath ${I}$}$_{n}$}%
+\phi_0^{-1} \mbox{{\boldmath ${K}$}$_{in}$}|+c +n\varepsilon  \notag \\
&\hskip 2cm -\log \int p_{0}(\mbox{{\boldmath ${y}$}$_{i}$}|r,%
\mbox{{\boldmath ${X}$}$_{i}$}) \exp\{-( \frac{r}{2}(||f_{0i}||^2_k+c))\} g(r)
dr.
\end{align}
By simple computation, we show that
\begin{align}  \label{gtilde}
&\int p_{0}(\mbox{{\boldmath ${y}$}$_{i}$}|r,\mbox{{\boldmath ${X}$}$_{i}$})
\exp\{-( \frac{r}{2}(||f_{0i}||^2_k+c))\} g(r) dr  \notag \\
=&\int p_{0}(\mbox{{\boldmath ${y}$}$_{i}$}|r,\mbox{{\boldmath ${X}$}$_{i}$}%
) g(r) dr \int \exp\{-( \frac{r}{2}(||f_{0i}||^2_k+c))\} {g}^*(r) dr,
\end{align}
where ${g}^*(r)$ is the density function of $IG(\nu+n/2,(%
\nu-1)+q_i^2/2)$. From (\ref{tpr1}), (\ref{tpr0}), (\ref{loggp}) and (\ref
{gtilde}), we have
\begin{align}
&-\log p_{\phi_0,\hat{\theta}}(\mbox{{\boldmath
${y}$}$_{i}$}|\mbox{{\boldmath ${X}$}$_{i}$})+\log p_{\phi_0}( %
\mbox{{\boldmath ${y}$}$_{i}$}|f_{0i},\mbox{{\boldmath ${X}$}$_{i}$})  \notag \\
\leq& \frac{1}{2}\log|\mbox{{\boldmath ${I}$}$_{n}$}+\phi_0^{-1} %
\mbox{{\boldmath ${K}$}$_{in}$}|+c-\log \int \exp\{-( \frac{r}{2}%
(||f_{0i}||^2_k+c))\} {g}^*(r) dr  \notag \\
\leq &\frac{1}{2}\log|\mbox{{\boldmath ${I}$}$_{n}$}+\phi_0^{-1} %
\mbox{{\boldmath ${K}$}$_{in}$}|+c+\frac{||f_{0i}||^2_k+c}{2} \int r {g}^*%
(r) dr  \notag \\
=&\frac{1}{2}\log|\mbox{{\boldmath ${I}$}$_{n}$}+\phi_0^{-1}
\mbox{{\boldmath
${K}$}$_{in}$}|+ \frac{q_i^2+2(\nu-1)}{2(n+2\nu-2)}(||f_{0i}||^2_k+c)+c+n%
\varepsilon,  \notag
\end{align}
which shows that Lemma 3 holds.

Now let us prove the inequality (\ref{GPcons}). Since the proof of (\ref
{GPcons}) is similar to those of Theorem 1 in \cite{r14} and Lemma
1 in \cite{r17}, here we summarily present the procedure of the
proof, details please see in \cite{r14}  and \cite{r17}.
Let $\mathcal{H}$ be the reproducing kernel Hilbert space (RKHS) associated
with covariance function $k_i(\cdot,\cdot;\vesub{\theta}{i})$, and $%
\mathcal{H}_n= \{\tilde{f}(\cdot): \tilde{f}(\cdot)=\sum_{l=1}^n\alpha_l k_i(%
\mbox{\boldmath ${x}$},\mbox{{\boldmath ${x}$}$_{il}$};\vesub{\theta}{i})$, for any $\alpha_l\in R\}$. From the Representer Theorem (see
Lemma 2 in Seeger \textit{et al}., 2008), it is sufficient to prove (\ref{GPcons})
for the true underlying function $\tilde{f}_{0i}=f_{0i}|r_i\in \mathcal{H}_n$. Then
for given $r_i$, ${f}_{0i} $ can be written as
\begin{align}
&{f}_{0i}(\cdot)=r_i\sum_{l=1}^n\alpha_l k_i(\mbox{\boldmath ${x}$},%
\mbox{{\boldmath ${x}$}$_{il}$};\vesub{\theta}{i})\doteq r_i K_i(\cdot)%
\mbox{\boldmath ${\alpha}$},  \notag
\end{align}
where $K_i(\cdot)=(k_i(\mbox{\boldmath ${x}$},\mbox{{\boldmath ${x}$}$_{i1}$};\vesub{\theta}{i}),...,k_i(\mbox{\boldmath ${x}$},%
\mbox{{\boldmath
${x}$}$_{in}$};\vesub{\theta}{i}))$ and $\mbox{\boldmath ${\alpha}$}%
=(\alpha_1,...,\alpha_n)^T$.

By Fenchel-Legendre duality relationship, we have
\begin{align}  \label{FLD}
-\log p_{G}(\mbox{{\boldmath ${y}$}$_{i}$}|r_i,\mbox{{\boldmath ${X}$}$_{i}$})
\leq E_{Q}(-\log {p}(\mbox{{\boldmath ${y}$}$_{i}$}|\tilde{f}_i,r_i))+D[Q,P],
\end{align}
where $P$ is a measure induced by $GP(0,r_ik_i(\cdot,\cdot;\hat{\ve\theta}_i))$, and $Q$ is the posterior distribution of $\tilde f_i$
from a GP model with prior $GP(0, r_ik_i(\cdot,\cdot;\vesub{\theta}{i}
))$ and Gaussian likelihood term $\prod_{l=1}^n N(\hat y_{il}|\tilde f_i(%
\mbox{{\boldmath ${x}$}$_{il}$}),r_i\phi_0)$, where $\hat{\ve{y}}_i=(\hat
y_{i1},...,\hat y_{in})^T=r_i( \mbox{{\boldmath ${K}$}$_{in}$}+\phi_0%
\mbox{{\boldmath
${I}$}$_{n}$})\mbox{\boldmath ${\alpha}$}$ and $%
\mbox{{\boldmath
${K}$}$_{in}$}=(k_i(\mbox{{\boldmath ${x}$}$_{ij}$},%
\mbox{{\boldmath
${x}$}$_{il}$};\vesub{\theta}{i}))_{n\times n}$. Then we have $E_{Q}(\tilde
f_i)=f_{0i}$, where the expectation is taken under probability density $Q$. Let $\mbox{\boldmath ${B}$}=\mbox{{\boldmath ${I}$}$_{n}$}%
+\phi_0^{-1} \mbox{{\boldmath ${K}$}$_{in}$}$, then we have
\begin{align}
&D[Q,P]=\frac{1}{2}\left\{-\log|\hat{\mbox{\boldmath ${K}$}}_{in}^{-1} %
\mbox{{\boldmath ${K}$}$_{in}$}|+\log|\mbox{\boldmath ${B}$}|+Tr(\hat{%
\mbox{\boldmath ${K}$}}_{in}^{-1} \mbox{{\boldmath ${K}$}$_{in}$}%
\mbox{{\boldmath ${B}$}$^{-1}$})\right.  \notag \\
&\hskip 2cm \left.+r_i||f_{0i}||^2_k+r_i\mbox{\boldmath ${\alpha }$} %
\mbox{{\boldmath ${K}$}$_{in}$}(\hat{\mbox{\boldmath ${K}$}}_{in}^{-1} %
\mbox{{\boldmath ${K}$}$_{in}$}-\mbox{{\boldmath ${I}$}$_{n}$})%
\mbox{\boldmath ${\alpha}$}-n\right\},  \label{KBdist} \\
&E_{Q}(-\log {p}(\mbox{{\boldmath ${y}$}$_{i}$}|\tilde{f}_i,r_i))\leq -\log
{p}(\mbox{{\boldmath ${y}$}$_{i}$}|f_{0i},r_i)+\frac{1}{2} \phi_0^{-1}Tr( %
\mbox{{\boldmath ${K}$}$_{in}$}\mbox{{\boldmath ${B}$}$^{-1}$})  \notag \\
&\hskip 3.2cm =-\log p_{0}(\mbox{{\boldmath ${y}$}$_{i}$}|r_i,%
\mbox{{\boldmath
${X}$}$_{i}$})+\frac{1}{2} \phi_0^{-1}Tr( \mbox{{\boldmath ${K}$}$_{in}$}%
\mbox{{\boldmath ${B}$}$^{-1}$}),  \label{EQP}
\end{align}
where $\hat{\mbox{\boldmath ${K}$}}_{in}=(k_i(\mbox{{\boldmath ${x}$}$_{ij}$},%
\mbox{{\boldmath ${x}$}$_{il}$};\hat{\mbox{\boldmath ${\theta}$}}_i))_{n\times n}$.

Hence, it follows from (\ref{FLD}), (\ref{KBdist}) and (\ref{EQP}) that
\begin{align}  \label{GPcons-1}
&-\log p_{G}(\mbox{{\boldmath ${y}$}$_{i}$}|r_i,\mbox{{\boldmath ${X}$}$_{i}$}%
)+\log p_{0}(\mbox{{\boldmath ${y}$}$_{i}$}|r_i,\mbox{{\boldmath ${X}$}$_{i}$})
\notag \\
\leq& \frac{1}{2}\left\{-\log|\hat{\mbox{\boldmath ${K}$}}_{in}^{-1} %
\mbox{{\boldmath ${K}$}$_{in}$}|+\log|\mbox{\boldmath ${B}$}|+Tr((\hat{%
\mbox{\boldmath ${K}$}}_{in}^{-1} \mbox{{\boldmath ${K}$}$_{in}$}+\phi_0^{-1} %
\mbox{{\boldmath ${K}$}$_{in}$})\mbox{{\boldmath ${B}$}$^{-1}$})
+r_i||f_{0i}||^2_k\right.  \notag \\
&\left.+r_i\mbox{\boldmath ${\alpha }$} \mbox{{\boldmath ${K}$}$_{in}$}(\hat{%
\mbox{\boldmath ${K}$}}_{in}^{-1} \mbox{{\boldmath ${K}$}$_{in}$}-%
\mbox{{\boldmath ${I}$}$_{n}$})\mbox{\boldmath ${\alpha}$}-n\right\}.
\end{align}
Since the covariance function is bounded and continuous in $%
\mbox{\boldmath
${\theta}$}_i$ and $\hat{\mbox{\boldmath ${\theta}$}}_i\rightarrow %
\mbox{\boldmath ${\theta}$}_i$, we have $\hat{\mbox{\boldmath ${K}$}}_{in}^{-1} %
\mbox{{\boldmath ${K}$}$_{in}$} -\mbox{{\boldmath ${I}$}$_{n}$}\rightarrow 0$
as $n\rightarrow \infty$. Hence, there exist positive constants $c$ and $%
\varepsilon$ such that for $n$ large enough
\begin{align}  \label{eqns}
&-\log|\hat{\mbox{\boldmath ${K}$}}_{in}^{-1} \mbox{{\boldmath ${K}$}$_{in}$}%
|<c,~~\mbox{\boldmath ${\alpha }$} \mbox{{\boldmath ${K}$}$_{in}$}(\hat{%
\mbox{\boldmath ${K}$}}_{in}^{-1} \mbox{{\boldmath ${K}$}$_{in}$}-%
\mbox{{\boldmath ${I}$}$_{n}$})\mbox{\boldmath ${\alpha}$}<c,  \notag \\
&Tr(\hat{\mbox{\boldmath ${K}$}}_{in}^{-1} \mbox{{\boldmath ${K}$}$_{in}$} %
\mbox{{\boldmath ${B}$}$^{-1}$})<Tr((\mbox{{\boldmath ${I}$}$_{n}$}%
+\varepsilon \mbox{{\boldmath ${K}$}$_{in}$})\mbox{{\boldmath ${B}$}$^{-1}$}).
\end{align}
Plugging (\ref{eqns}) in (\ref{GPcons-1}), we have the inequality (\ref
{GPcons}). $\sharp$

\vskip 10pt

To prove Proposition 5, we need condition \newline
(A) $||f_{0i}||_{k}$ is bounded and $E_{\mbox{{\boldmath
${X}$}$_{i}$}}(\log|\mbox{{\boldmath ${I}$}$_{n}$}+\phi_0^{-1}%
\mbox{{\boldmath
${K}$}$_{in}$}|)=o(n)$.

\vskip10pt \noindent \textbf{Proof of Proposition 5}: It
easily shows that $q_i^2=(\mbox{{\boldmath ${y}$}$_{i}$}-f_{0i}(\mbox{{\boldmath
${X}$}$_{i}$}))^T(\mbox{{\boldmath ${y}$}$_{i}$}-f_{0i}(\mbox{{\boldmath
${X}$}$_{i}$}))/\phi_0=O(n) $. Under conditions in Lemma 3, and condition
(A), it follows from Lemma 3 that
\begin{equation}
\frac{1}{n}E_{\mbox{{\boldmath ${X}$}$_{i}$}}(D[p_{\phi _{0}}(%
\mbox{{\boldmath
${y}$}$_{i}$}|f_{0i},\mbox{{\boldmath ${X}$}$_{i}$}),p_{\phi _{0},\hat{\theta}_i%
}(\mbox{{\boldmath ${y}$}$_{i}$}|\mbox{{\boldmath ${X}$}$_{i}$}%
)])\longrightarrow 0,{\mbox{as}}~~n\rightarrow \infty.  \notag
\end{equation}
Hence, Proposition 3 holds.$\sharp$

\vskip 10pt
\section{{  More simulation studies}}

\begin{figure}[h]
\begin{center}
\includegraphics[height = 0.7\textwidth,width=0.98\textwidth]{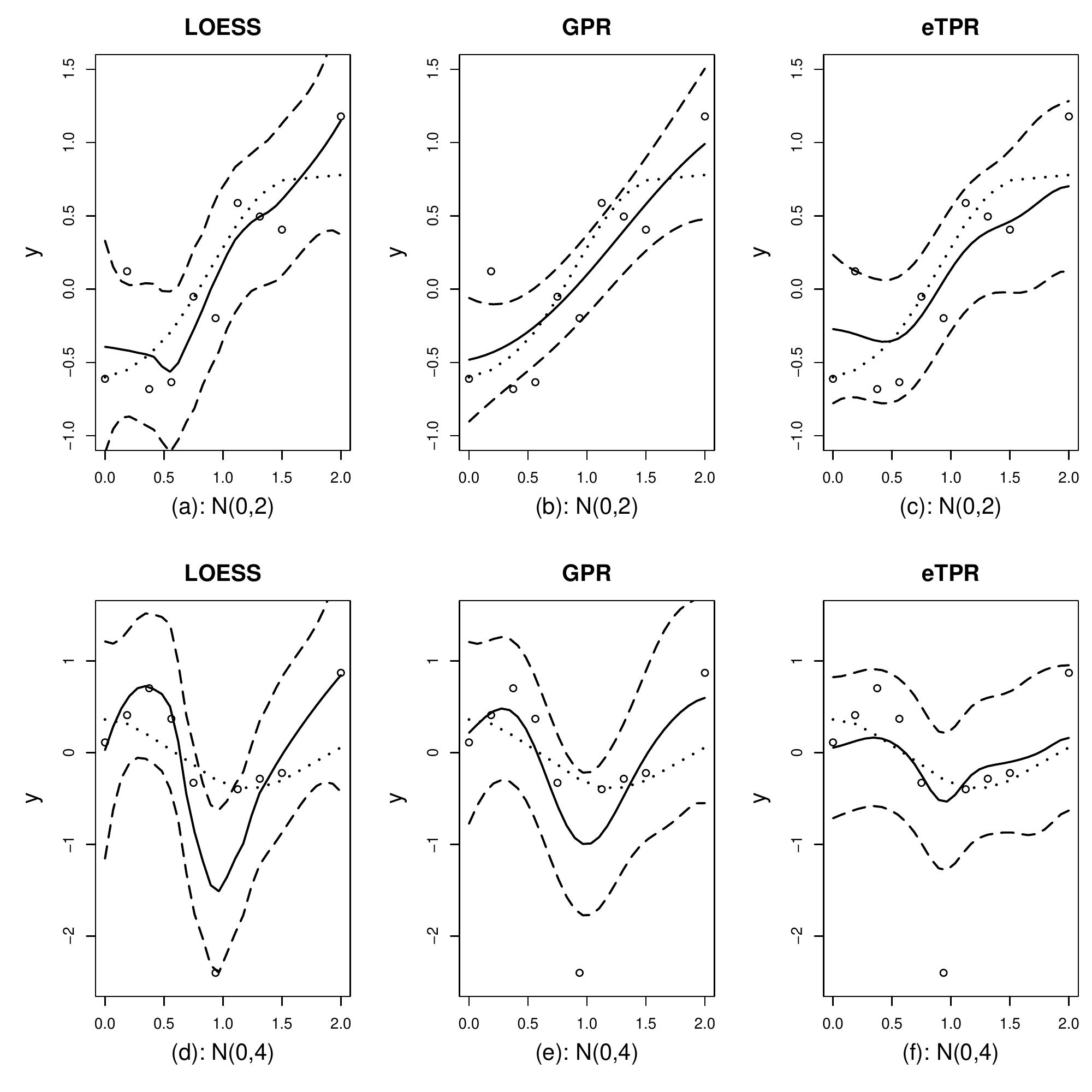}
\end{center}
\caption{Predictions in the presence of outlier at the middle data point which is disturbed
by additional error generated from $N(0,\protect\sigma^2)$, where circles represent the observed
data, dotted line is the true function, and solid and dashed lines
stand for predicted curves and their 95\% point-wise confidence intervals from the LOESS,
GPR and eTPR methods respectively. }
\label{figD1}
\end{figure}

\begin{figure}[h]
\begin{center}
\includegraphics[height = 0.7\textwidth,width=0.98\textwidth]{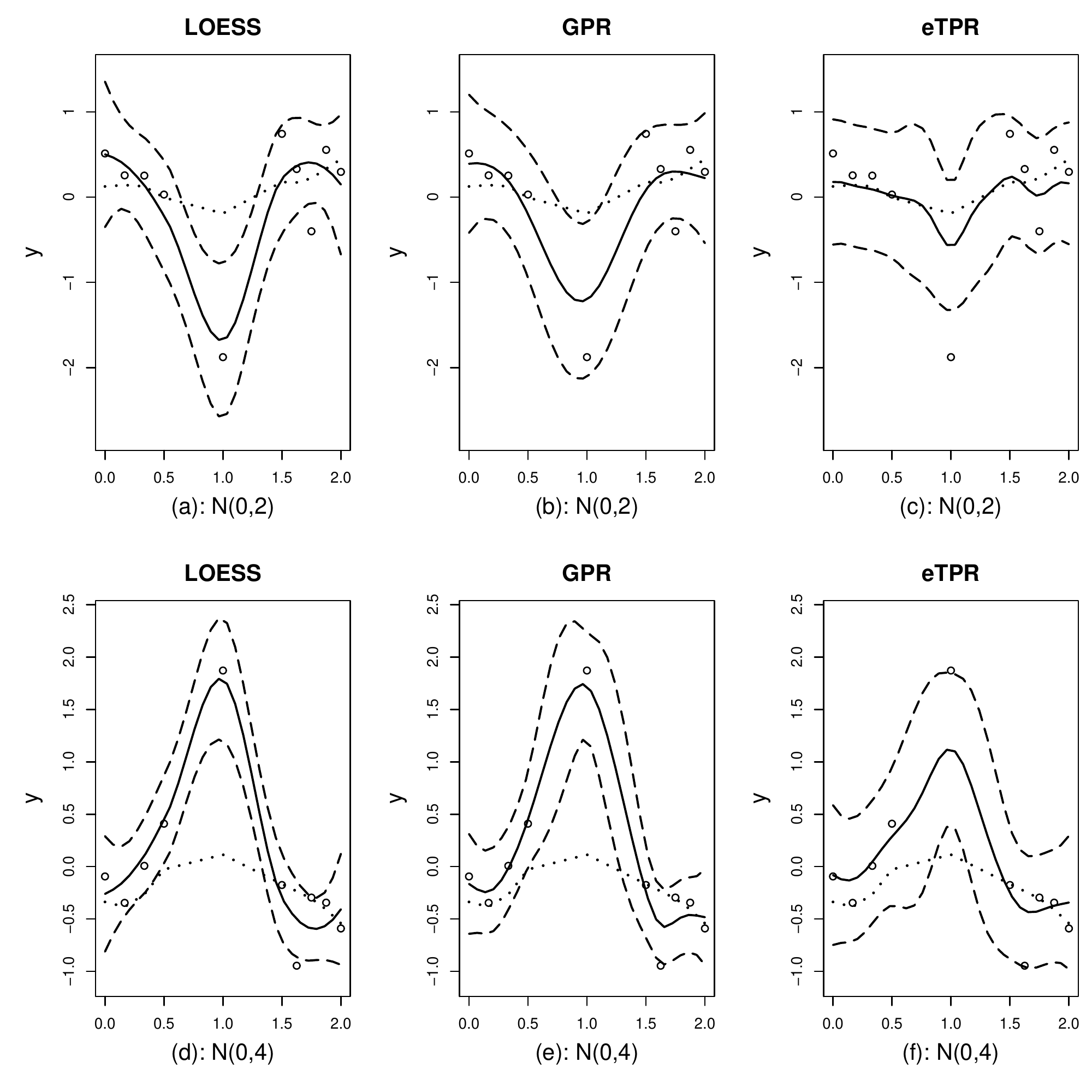}
\end{center}
\caption{Predictions in the presence of sparse and outlier at middle point 1.0 which is disturbed
by additional error generated from $N(0,\protect\sigma^2)$, where circles represent the observed
data, dotted line is the true function, and solid and dashed lines
stand for predicted curves and their 95\% point-wise confidence intervals from the LOESS,
GPR and eTPR methods respectively. }
\label{figD2}
\end{figure}
{ 
Two data sets with $m=1$ and sample
size of $n_1=10$ are generated, where the first $n_1-1$ data points $\{x_{1j}\}$ are evenly spaced in [0,~1.5] and the remaining point is at 2.0.
At the middle point of $x_{1j}=0.9375$, the observation is added with an extra error from either $N(0,2)$ or $
N(0,4)$. Prediction curves,  the observed data, the true
function and their 95\%
point-wise confidence intervals, are plotted in Figure \ref
{figD1}. We see that the  predictions from eTPR shrinks heavily in the area near the data point 1.0, compared to LOESS and GPR.

In addition, we make the data sparse in the area near the point 1.0  by  generating the first $4$ and the last $5$ data points $\{x_{1j}\}$ evenly spaced in [0,~0.5]
and [1.5,~2.0], respectively, and to make it outlier by adding an extra error from either $N(0,2)$ or $
N(0,4)$. Other setups are the same as those in Figure \ref{fig1}. Results are presented in Figure \ref
{figD2}. It shows that the predictions from eTPR also shrinks heavily and have more robustness around the data point 1.0, compared with LOESS and GPR.

\begin{table}[!ht]
\caption{Mean squared errors of prediction results and their standard
deviation (in parentheses) by the LOESS, GPR and eTPR methods with $n=60$ and 100.}
\label{Dtab1}\tabcolsep=5pt \fontsize{10}{16}\selectfont
 \vskip 0pt
\par
\begin{center}
\begin{tabular}{ccccc} \hline
n&Model & LOESS & GPR & eTPR \\ \hline
60&(1)&0.124(0.806)&0.025(0.063)&0.019(0.042)\\
&(2)&0.127(0.803)&0.032(0.079)&0.028(0.076)\\
&(3)&0.063(0.171)&0.036(0.046)&0.032(0.042)\\
&(4)&0.066(0.172)&0.048(0.061)&0.045(0.071)\\
&(5)&0.101(0.550)&0.032(0.086)&0.027(0.060)\\
&(6)&0.346(1.016)&0.264(0.551)&0.268(0.598)\\

100&(1)&0.152(1.978)&0.017(0.059)&0.013(0.036)\\
&(2)&0.156(1.980)&0.024(0.082)&0.022(0.071)\\
&(3)&0.038(0.102)&0.026(0.032)&0.023(0.025)\\
&(4)&0.041(0.103)&0.036(0.040)&0.033(0.035)\\
&(5)&0.174(2.220)&0.018(0.049)&0.015(0.027)\\
&(6)&0.274(0.719)&0.215(0.349)&0.211(0.346)\\

\hline
\end{tabular}
\end{center}
\end{table}

%
%

To illustrate performance of eTPR with large sample size, data $y_{1j}$ with $n=60$ and 100 are generated from the 6 process models in part (i) in Subsection 5.1. 
Other setups are the same as those in Table \ref{tab2}, where $N=100$ (150)\footnote{100 or 150?}.
From Table \ref{Dtab1}, we see that the eTPR method performs better or comparable with GPR, and both are better than LOESS. Combined with the results shown in Table 3, eTPR performs overall better than GPR, but the performance tends to similar when the sample size increases. This matches     the theory presented in Proposition 1 that the extended T-process behaves similar to Gaussian process when sample size $n$ is large.

}

\section*{Acknowledgements}

Wang's work is supported by funds of the State Key Program of National Natural Science of China (No. 11231010)  and National Natural Science of China (No. 11471302).

\end{document}